\documentclass{article}
\usepackage{graphicx} 

\title{What Does `(Non)-Absoluteness of Observed Events’ Mean?}
\author{Emily Adlam  \thanks{Philosophy Department, Chapman University, Orange, CA92866, USA; The Rotman Institute of Philosophy, 1151 Richmond Street, London N6A5B7 \texttt{eadlam90@gmail.com} }}
\date{\today}

\begin{document}

\maketitle

\begin{abstract}

Recently there have emerged an assortment of  theorems relating to the `absoluteness of emerged events,' and   these results have sometimes been used to argue that    quantum mechanics may involve some kind of metaphysically radical non-absoluteness, such as relationalism or perspectivalism. However,   in our view a close examination of these theorems fails to convincingly support such possibilities. In this paper we argue that the Wigner's friend paradox, the theorem of Bong et al and the theorem of Lawrence et al are all best understood as demonstrating that if quantum mechanics is universal, and if certain auxiliary assumptions hold, then the world inevitably includes various forms of   `disaccord,' but this need not be interpreted in a metaphysically radical way; meanwhile, the theorem of Ormrod and Barrett is best understood either as an argument for an interpretation allowing multiple outcomes per observer, such as the Everett approach, or as a proof that quantum mechanics cannot be universal in the sense relevant for this theorem. We also argue that these theorems taken together suggest interesting possibilities for a different kind of relational approach in which  \emph{dynamical} states are   relativized  whilst observed events are absolute, and we show that although something like `retrocausality'  might be needed to make such an approach work, this would be a very special kind of retrocausality which would evade a number of common objections against retrocausality. We conclude that the non-absoluteness theorems may have a significant role to play in helping converge towards an acceptable solution to the measurement problem.

\end{abstract}

\newpage

\section{Introduction}

Recently there have emerged an assortment of  theorems relating to the `absoluteness of emerged events.' Various interpretations of these results are possible, but one influential school of thought suggests these theorems demonstrate that if certain kinds of experiments were to have the results predicted by unitary quantum mechanics, then we would most likely have to accept that even observed events are not `absolute'\footnote{We emphasize that this response is not necessarily endorsed by the original authors of the theorems, who in many cases adopt an attitude to neutrality towards the various possible ways one could respond to their theorem.}\cite{2022NatRP...4..628B,mbitbol,subjectivefacts}. For example, it has been suggested that this could involve a picture in which measurements only have definite outcomes relative to individual observers  - i.e. there is no absolute, `third-person’ view from which we can say which outcome a measurement actually had\cite{Oldofredi2022-OLDTRD,Pienaar_2021}. 
 
Now, clearly one way in which measurements could be `non-absolute' in this way would be if a given observer sometimes observes more than one outcome for a given measurement, as for example in the Everett interpretation or other multiple-outcome-per-observer (MOPO) approaches. But the  non-absoluteness theorems have sometimes been used to argue for some \emph{other} kind of non-absoluteness  - in particular, for a version of relationalism or perspectivalism in which there is only one outcome per observer per measurement, but different observers may disagree about the outcome of a given measurement. However, in our view a close examination of these theorems fails to convincingly support such a possibility. In this paper we will argue that  the Wigner's friend paradox\cite{wigner1995remarks}, the Bong et al theorem\cite{Bong_2020} and the Lawrence et al theorem\cite{Lawrence_2023} do indeed demonstrate that if quantum mechanics is universal, and if certain auxiliary assumptions hold, then the world inevitably includes various forms of   `disaccord,' by which we mean   circumstances in which observers may reasonably fail to agree about the outcome of a given measurement - but there is no compelling reason to interpret this `disaccord' in terms of metaphysically radical forms of relationalism or perspectivalism. Meanwhile, we find that the theorem of Ormrod and Barrett\cite{ormrod2022nogo} could be regarded as offering a genuine argument for metaphysically radical non-absoluteness, but this is achieved only by making an unusually strong assumption about the circumstances in which unitary quantum mechanics makes correct predictions, and our view is that the theorem should be interpreted either as an argument in favour of MOPO approaches or simply as a reductio ad absurdum against this assumption.

However, this does not mean that the non-absoluteness theorems are not useful; on the contrary, we believe that the emphasis on metaphysically radical interpretations of these theorems may be obscuring some very important lessons that could be drawn from them. In particular, we argue that the Wigner's friend paradox could be regarded as demonstrating that if quantum mechanics is universal then \emph{dynamical states}  must be relativized, even though the events that actually occur are absolute. We note that the non-absoluteness theorems make it clear that such an approach would either have to violate Locality or would have to exhibit something like superdeterminism or retrocausality; but they also help us see that  the kind of `retrocausality' required is of a very special kind, such that common objections against retrocausality may not apply to it. We also demonstrate that this vision of a dynamically relational, retrocausal approach to quantum mechanics is already realised by seversal existing interpretations.  We conclude that the non-absoluteness theorems have significantly narrowed  the space of viable interpretations of quantum mechanics by demonstrating that workable approaches must have some quite specific properties, so they may play a significant role in helping us converge towards an acceptable solution to the measurement problem.

\section{Background \label{Intro}}

The non-absoluteness theorems are descendants of the Wigner’s friend paradox\cite{wigner1995remarks}, which describes a scenario in which some observer, Chidi, performs a measurement $\{ |S_i \rangle \langle S_i | \}$ on a system $S$, and then another observer, Alice, performs a measurement on the joint system of Chidi and $S$. The supposed paradox is that Chidi will presumably have seen a definite outcome to his measurement, so he will ascribe some state $| S_i \rangle$ to the system $S$, and yet if we believe that quantum mechanics is universal then the correct way to describe the interaction between Chidi and $S$ is to say that they end up in a state $\psi = \sum_i c_i |C_i \rangle |S_i \rangle$, where $|C_i \rangle$ is the state of Chidi corresponding to him having seen the outcome $| S_i\rangle$ to his measurement. Moreover, if this experiment is repeated many times Alice can in principle confirm that Chidi and $S$ do end up in this state, provided that she is able to maintain complete coherent control of the joint Chidi-$S$ system long enough to perform tomographic measurements.   Yet the state $\psi$ appears to represent Chidi as not obtaining any definite outcome to his measurement, so how can it be the case that Chidi has observed a definite outcome if Alice subsequently finds him in the state $\psi$ ?

As just stated this `paradox’ is perhaps puzzling but does not lead to an outright contradiction, so the class of theorems that we will refer to as the `non-absoluteness theorems’ have set out to provide something more watertight using `extended Wigner's friend' scenarios. The first such theorem that we will discuss is  due  to Bong et al\cite{Bong_2020}. It should be emphasized Bong et al originally derived their theorem in a theory-independent way, arriving at some inequalities which must be obeyed by any theory satisfying their assumptions Absoluteness of Observed Events, Locality and No-Superdeterminism.  But in this paper we are specifically interested in understanding whether Bong et al's theorem can be used to argue that the universality of quantum mechanics would entail the existence of some kind of non-absoluteness, so we will follow the presentation of ref  \cite{schmid2023review} and work directly in the context of quantum theory. Thus suppose we have two agents, Chidi and Divya, each in a closed laboratory, and each in possession of one particle from an entangled  pair.  Chidi and Divya each perform a measurement of a certain fixed observable of their particle, obtaining measurement results C, D. Then we have another observer Alice who performs a `supermeasurement'  on the whole system of Chidi’s closed lab, obtaining an outcome A; and a fourth observer, Bob, who performs a supermeasurement  on Divya's closed lab, obtaining an outcome B\footnote{Ref \cite{Bong_2020} also includes the more general case in which Alice and Bob can choose between two or more supermeasurements, which is useful to show that the set of correlations obeying the inequalities derived from their assumptions is strictly larger than the set obeying inequalities obtained just from Bell's locality assumptions. However, only one supermeasurement per observer is needed to derive the `non-absoluteness' result we are interested in here, so we will focus on that simple case.}.  The word `supermeasurement' here just refers to the fact that one observer is  performing a measurement on another observer, and that this is being done in a basis which does not commute with the variables encoding that observer's memories, so for example here Alice is performing on Chidi a measurement which does not commute with the variable encoding the result of Chidi's measurement C, and likewise for Bob and Divya. We use this terminology to emphasize that in practice, performing  such a  measurement requires Alice to maintain complete coherent control over Chidi and his system, which means she must ensure that no information whatsoever escapes Chidi's closed laboratory, and she must be able to exert fine control over each individual degree of freedom making up Chidi and his system.

 Bong et al argue that `absoluteness of observed events’ (AOE) means, first of all, that there are  is exactly one outcome of each measurement per observer, so in this case we have exactly four measurement outcomes  witnessed by the person who performs the relevant experiment -  $A, B, C$ and $D$. In addition, Bong et al state that AOE also entails that if Alice, instead of performing a supermeasurement, chooses to perform a measurement which tells her the result of Chidi’s measurement (e.g. by simply going into the laboratory and asking Chidi what result he got) then the outcome of this measurement must match Chidi’s outcome C; and likewise mutatis mutandis for Bob and Divya. Bong et al also invoke assumptions that they call Locality and No-Superdeterminism to argue that the values of C and D cannot depend on whether Alice and Bob choose to perform their supermeasurements or to ask Chidi and Divya what results they obtained, so we have a fixed set of four outcomes $A, B, C, D$ regardless of what measurements Alice and Bob perform. Now,  let us suppose  that after the experiment Bob shares his result with Alice; thus a single observer, Alice, could ultimately come to know the values of any one of the pairs in the set $\{ AB, AD,  CB,   CD\}$, drawn from the fixed set of outcomes $A, B, C, D$. Then since we have chosen to work directly in the context of quantum theory, we may proceed on the basis of the assumption  that quantum mechanics is universal, in the sense that  Alice will never see an outcome which is in contradiction with unitary quantum mechanics; it then follows that in any run of this experiment the four fixed outcomes  must match the predictions made by quantum mechanics for each of the pairs of variables in the set $\{ AB, AD,  CB,   CD\}$.  However, it is possible to choose the states and measurements used in this experiment such that   there is no possible choice for the values of $A, B, C$ and $D$   which  reproduces the predictions of unitary quantum mechanics for all of these pairs. Thus if it is really the case  that unitary quantum mechanics is universal,  and we are not willing to deny Locality or No-Superdeterminism, it appears we must accept that observed events are not absolute, in the sense in which that term is used by Bong et al\footnote{Supermeasurements on real observers are not possible using current technology, and most likely will never be possible using any realistic technology, so this experiment is and will likely remain a thought experiment. However,  a later version of this theorem\cite{wiseman2023thoughtful} considers the possibility of replacing Chidi with something like an artificial intelligence, focusing on the conditions that would need to be met in order for us to agree that such a device has genuinely made an observation which we would naturally expect to be `absolute.' This version of the experiment might well be performable at some time in the future. However, in this paper we will focus on the original Bong et al theorem, because as long as one is convinced that a real observation can be made by the   relevant device in this experiment, its foundational consequences seem  roughly the same as the consequences of the original experiment. One exception to this  is that the two experiments may possibly have different consequences for approaches suggesting that the interpretation of quantum mechanics is linked in some intrinsic way to (human) consciousness, but we are not considering  such possibilities here, and thus  we think most of our analysis in this paper will also apply to the newer theorem.}.

Another such theorem was proved by Lawrence et al\cite{Lawrence_2023}. It uses a scenario similar to the Bong et al experiment, except that it employs GHZ states rather than Bell pairs, and there are only two agents: first Alice measures all three GHZ qubits, each in a fixed basis, and then Bob performs three different supermeasurements on the joint system composed of Alice and her three qubits, each in a fixed basis, with each of Bob’s measurements accessing the Hilbert space associated linearly with one of the original qubits. Thus Lawrence et al argue that if there are `relative facts' about the outcomes of these measurements relative to Alice and Bob respectively, this experiment must produce exactly six measurement outcomes,  $A_1, A_2, A_3; B_1, B_2, B_3$, each relativized to the person performing the relevant measurement. Lawrence et al  contend that  these outcomes should obey certain constraints imposed by quantum mechanics: for example, they show that quantum mechanics predicts  that with an appropriate labeling convention, the product of the values $B_1 B_2 B_3$ will always be equal   to $1$, and similar constraints can be obtained for the trios $ B_1 A_2 A_3$, $A_1 B_2 A_3$, and $A_1 A_2 B_3$.  But it is straightforward to show that there is no  possible assignation of values to  $A_1, A_2, A_3; B_1, B_2, B_3$ which obeys all four of these constraints, and thus Lawrence et al argue that `relative facts' are inconsistent with quantum mechanics. 

Although Lawrence et al are concerned with relative facts rather than absoluteness of observed outcomes here, their theorem can be rewritten so as to follow the same argument pattern as that of Bong et al. One could use AOE rather than the existence of relative facts to argue that there are exactly six measurement outcomes  witnessed by the person who performs the relevant experiment; and second, one could follow the reasoning of Bong et al to argue that if Bob performs a measurement on Alice seeking to learn her measurement outcomes, he must learn the actual values of $A_1, A_2$ and/or $A_3$. Thus one could conclude that Bob is principle able to come to know the values of any one of the trios $\{ B_1 B_2 B_3, B_1 A_2 A_3, A_1 B_2 A_3, A_1 A_2 B_3 \}$\footnote{Obviously Bob could also come to know $B_1 B_2 A_3, B_1 A_2 B_3, A_1 B_2 B_3$ or $A_1 A_2 A_3$, but these trios are not needed for the proof.}, and then one could invoke the universality of quantum mechanics in the first-person sense to argue that on any run of the experiment the values of the outcomes $A_1, A_2, A_3; B_1, B_2, B_3$ must obey the constraints imposed   by quantum mechanics for all four trios, which is known to be impossible.  Thus the Lawrence et al argument could be used just like the Bong et al argument to  argue that the universality of quantum mechanics implies that observed events are not absolute. Note that  Lawrence et al do not explicitly mention assumptions about superdeterminism and retrocausality,  but an examination of their derivation makes it clear that they are in fact making an implicit assumption very similar to Bong et al's No-Superdeterminism assumption,  since they take it that the quantum constraints must always be obeyed for all four trios, whereas if the values of $A_1, A_2, A_3$ were allowed to depend on what variables Bob chooses to measure  then the quantum constraints would only need to be obeyed for the one trio whose value Bob does actually come to know. The Lawrence et al result also requires an assumption of Locality, because the proof requires us to assume that when Bob measures $B_1$ the value of $B_1$ is independent of whether he measures $B_2$ or $A_2$, and so on mutatis mutandis; so to make this argument work we probably need to imagine that `Bob' is really several people spread out in space as in the Bong et al case, in order that we can  apply Locality to conclude that the results of these measurements are independent. 
 
Finally we have the theorem of Ormrod and Barrett\cite{ormrod2022nogo}, which uses a similar experimental setup to Bong et al. Ormrod and Barrett consider four different spacelike slices on which we could in principle collapse the wavefunction (one including the measurements of Alice+Bob, one including the measurements of Chidi+Divya, one including the measurements of Alice+Divya, and one including the measurement of Bob+Chidi), and show that if we try to make predictions for measurement outcomes in four different ways, by collapsing the wavefunctions on each of these four spacelike slices, then certain pairs of results on the Chidi-Divya, Bob-Chidi and Alice-Diya spacelike slices are assigned probability zero. Then if it is assumed that observed events are absolute in the sense that for each run of this experiment there exist exactly four non-relativized outcomes $A, B, C, D$, we can infer that a certain pair of results on the Alice-Bob spacelike slice must also be impossible - but unitary quantum mechanics predicts that this pair of  results \emph{is} possible, so if unitary quantum mechanics is universal this pair of results would presumably be seen if the experiment were repeated enough times.   Thus they conclude that if unitary quantum mechanics is universal in the sense that we can always choose to collapse the wavefunction on any arbitrary spacelike slice, then observed events cannot be absolute. The theorem of Ormrod and Barrett requires neither a Locality assumption nor a No-Superdeterminism (and/or retrocausality) assumption.

This does not exhaust the space of non-absoluteness theorems in the literature - we will not be able to cover them all in this paper, but ref \cite{schmid2023review} provides a helpful overview of some other non-absoluteness theorems, including explanations of how they relate to the theorems discussed here. In our view, none of these other theorems convincingly support the possibility of metaphysically radical non-absoluteness either, for reasons similar to the ones we will shortly set out with regard to the Bong et al, Lawrence et al and Ormrod and Barrett theorems.

\subsection{Universality}

Now, we emphasize that none of the experiments referenced in the non-absoluteness theorems have yet been performed, and there is no prospect that they will be performed in the immediate future. So it is not possible to use these theorems to argue directly for the non-absoluteness of observed events on the basis of real empirical results. Rather it is necessary to begin by making some assumption about what the results of these experiments will be - and as noted above, typically those who wish to use the theorems to argue for non-absoluteness  proceed by assuming `the universality of unitary quantum mechanics,' so it can be assumed the results of these experiments will be as predicted by unitary quantum mechanics\footnote{We note that a more recent version of the Bong et al theorem, in ref \cite{wiseman2023thoughtful}, assumes only that universal quantum computing is valid; and more generally, it is clear that to make these theorems work one need only assume that the predictions of unitary quantum mechanics are correct for one particular experiment, not that they are always correct. However, presumably `the universality of unitary quantum mechanics' implies at least that its predictions are correct for this particular experiment, so the reasons we adduce in this section for believing in the universality of unitary quantum mechanics are automatically also reasons for believing any such weaker assumption.}. However,  there are different ways in which quantum mechanics could be `universal,' and therefore  in this paper we will need to keep in mind  two different conceptions of universality. The first is `first-person' universality: to say that unitary quantum mechanics is `universal' in this sense is to say that it is always a correct description of all the observations made by any individual observer, but it can't in general be applied jointly across the observations of several different observers. This is the kind of universality which must be assumed if we wish to use the theorems of Bong et al and Lawrence et al to argue for non-absoluteness, since those authors are at pains to ensure that the sets of variables used to prove their theorems are all in principle available to a single individual. First-person universality is  also the kind of universality which is relevant for  relational and perspectival interpretations of quantum mechanics\cite{FuchsMermin, 1996cr}, in which quantum mechanics is typically understood as a single-user, first-person theory designed to characterise observations from the point of view of the individual observer who is currently applying it. 

However, one might also think that unitary quantum mechanics should be universal in a stronger sense. For example, the theorem of Ormrod and Barrett uses a stronger notion of universality; motivated by relavistic considerations, they argue that   no spacelike slices should be privileged and thus the `universality of unitary quantum mechanics' should entail that we can always make correct predictions by choosing an arbitrary spacelike slice and then applying unitary evolution   up until a collapse of the wavefunction on that spacelike slice, with predictions on that slice obtained from the Born rule. They refer to this prescription as `Frame-Independent Quantum Theory' (FIQT). Evidently universality in the FIQT sense is a stronger requirement than the first-person approach to universality, since it involves applying quantum mechanics directly across observations made by different observers on the same spacelike slice, whereas the first-person approach would not recognise as legitimate applications of the theory across several different observers in this way. It must be emphasized that although this kind of universality may seem very reasonable, it can never be directly empirically verified, because ultimately the only thing we can ever directly verify is the fact that quantum mechanics makes the right predictions for an individual observer  (including predictions about what other observers will report to them about their own observations). And therefore, although the theorem of Ormrod and Barrett does not assume No-Superdeterminism or Locality and is thus in some sense stronger than the Bong et al and Lawrence et al theorems, in another sense it is weaker because it assumes a more demanding version of `the universality of unitary quantum mechanics.'

\section{Responses \label{resp}}

There  is one obvious route which allows us to accommodate the non-absoluteness theorems without accepting any kind of non-absoluteness  -  we need only adopt any interpretation of quantum mechanics which postulates something like a von Neumann cut, such that unitary quantum mechanics applies only in certain regimes and is therefore not universal in either of the two senses discussed above. For example, spontaneous collapse interpretations\cite{Frigg2009,Tumulka2006} and interpretations where the wavefunction is collapsed by consciousness\cite{london1939theorie} do not predict that the experiments referenced in the non-absoluteness theorems would have the results we described in section \ref{Intro} which lead to lead to contradictions with AOE, and therefore proponents of spontaneous collapse and consciousness-based collapse interpretations clearly do not need to worry about non-absoluteness. 

However, there are  good scientific reasons for being wary of approaches in which unitary quantum mechanics is not universal in at least one of these senses. For a start, we have not yet found any direct evidence that quantum mechanics ceases to apply in some regimes, so postulating a von Neumann cut involves adding a considerable amount of additional structure without any clear empirical basis. Additionally, Wallace\cite{https://doi.org/10.48550/arxiv.2205.00568} points out that no extant spontaneous collapse interpretation can reproduce all the predictions of quantum field theory, and he also adduces various structural reasons to think that other approaches involving something like a von Neumann cut are likely to face similar difficulties, so as things stand it seems possible that   only  approaches which uphold the universality of unitary quantum mechanics will ultimately be empirically adequate. Thus it is certainly tempting to respond to the non-absoluteness theorems by simply accepting that indeed, observed events are not absolute. As a physicist one may  feel this is a small cost to pay if the alternative requires us to deny the universality of unitary quantum mechanics and then do all the hard work of constructing what would essentially be a whole different scientific theory incorporating some second kind of dynamical process.

However, accepting the non-absoluteness of observed events is not as harmless as it may seem: in fact this move has very serious consequences for the epistemology of science. For example, one way of denying  AOE is to adopt an Everettian approach in which all of the possible outcomes for a given measurement do actually occur in different branches of the wavefunction - but as argued by ref \cite{AdlamEverett}, it is difficult in this kind of picture to give an account of probability adequate to make sense of probabilistic confirmation, which is surely essential to scientific practice. Another way of denying  AOE is to say that no measurement has more than one outcome relative to each observer but nonetheless different observers may fail to agree about the outcome of a given measurement - yet as argued by ref \cite{https://doi.org/10.48550/arxiv.2203.16278}, this may lead to a picture in which observers cannot ever share any information about measurement outcomes and thus it is impossible to use observations made by other observers or even past versions of oneself for the purpose of empirical confirmation, which likewise seems essential to scientific practice. 

Now, of course it could be that there exists some other kind of `non-absolute' approach which does not face the same kinds of problems, or perhaps there is some way to solve the problems we have just described for the  existing approaches.   For example, the Everettians have proposed a number of different ways of obtaining probabilities within their picture\cite{Greaves, Wallace,vaidman1996schizophrenic}, although it is still unclear whether any of them is suitable to address the empirical confirmation issue in particular. But at any rate, it is clear that accepting the non-absoluteness of observed events should not be done lightly - this move risks endangering our scientific methodology to the extent that it may no longer be possible to regard quantum mechanics as   empirically confirmed, and  we certainly cannot rationally believe an interpretation of quantum mechanics which denies that quantum mechanics has been empirically confirmed. Thus one might begin to worry that  these non-absoluteness theorems  will ultimately compel us to   accept that quantum mechanics cannot be universal in any sense. 

However, there may be another way out. This is because the terms `absolute’ and `relative’ are somewhat vague and abstract, and    discussions of these theorems often end up equivocating between different meanings of these words. So it may be that it is possible to come up with an approach which is `absolute’ enough to avoid the kinds of epistemic problems that we have just alluded to, while still being  `relational' enough to be compatible with the universality of quantum mechanics. In this article we will seek to understand what such an approach might look like.

\section{Non-Absoluteness \label{NA}}

Non-absoluteness theorems typically define `non-absoluteness' by negation: that is, they provide a definition for `absoluteness of observed events' and then `non-absoluteness of observed events' is  understood to refer to any scenario in which that definition fails to hold.    For example, both Bong et al and Ormrod and Barrett specify that AOE involves the stipulation that `\emph{an observed event is a real single event, not relative to anything or anyone}' and `non-absoluteness of observed events' is simply the denial of this. 

However,  if we are to use these theorems to argue that the universality of unitary quantum mechanics implies the `non-absoluteness of observed events' (conditional on some auxiliary assumptions), it is crucial to say something about what `non-absoluteness' might actually look like. For if non-absoluteness is merely defined as the negation of some notion of `absoluteness,' that leaves open the possibility that there just is not \emph{any coherent way}  in which events could fail to be absolute in this particular sense. And if so, the relevant non-absoluteness theorem has not proved the existence of non-absoluteness: rather, it has proved that, if we are not willing to let go of any of the auxiliary assumptions, then we must accept that quantum mechanics is not universal, i.e. the theorem has become a reductio ad absurdum argument against the universality of unitary quantum mechanics.

Now as matter of fact, we think there is at least one coherent way in which observed events could fail to be absolute: an approach which allows a given measurement to have multiple outcomes per observer (MOPO), such as the Everett interpretation. Clearly in an Everettian world  measurement outcomes are   not absolute - there is no fact of the matter about which outcome of a measurement actually occurred, given that  all of the possible outcomes do actually occur. But in this paper our aim is to investigate  possible alternatives to these multiple-outcome pictures, so we will henceforth imagine that all MOPO approaches  have been judged untenable - in which case, if there is no other coherent way in which observed events could fail to be absolute, and we are not willing to deny any auxiliary assumptions, we would indeed be forced to conclude that quantum mechanics is not universal. 

However, it is commonly suggested in the literature that there \emph{is} some other coherent way in which observed events could fail to be absolute. In particular, in `relational' or `perspectival' approaches\cite{Bohr1987-BOHTPW,Heisenberg1958-HEIPAP,pauli1994writings,Zeilinger1999-ZEIAFP,Zeilinger2002,brukner2015quantum,articlebanana,demopoulos2012generalized,Janas2021-JANUQR,2004neoc,QBismintro,1996cr}, no measurement has more than one outcome relative to any observer, but nonetheless there is no fact of the matter about which outcome of a measurement actually occurred because a measurement can have different outcomes relative to different observers. These kinds of views are typically associated with quite strong metaphysical claims - for example, proponents of such views have suggested that there is no objective reality\cite{subjectivefacts}; that all physical facts must be relativized to an observer\cite{brukner2015quantum}; that facts are subjective\cite{subjectivefacts}; that it is not even meaningful to compare the perspectives of different observers\cite{pittphilsci19664};   and so on. We will henceforth refer to these kinds of  possibilities   as  `metaphysically radical' non-absoluteness. (In this paper we will not include the Everett interpretation and other MOPO approaches in the category of metaphysically radical non-absoluteness). 

Now, one may naturally wonder how it can be that a measurement can have different outcomes relative to different observers. After all, in each of these approaches it is accepted that there is a specific observer who actually performs the measurement and witnesses exactly one outcome, and one might naturally think that the unique outcome of the measurement relative to the observer who actually performed that measurement should be regarded as a unique, `absolute’ fact about the outcome of the measurement. If Chidi is the person who actually performed the measurement, isn’t he the ultimate authority on his own experience? If Alice fails to agree with him about that outcome, why should we say that there is a different measurement outcome relative to her? Isn’t she just \emph{wrong}? 

Thus we think it is not particularly helpful to simply gesture at the possibility that observers may fail to agree about measurement outcomes; it is important to be clear  about what that actually means. And in fact, an examination of the  non-absoluteness theorems discussed in section \ref{Intro} reveals at least three different ways in which one observer may fail to agree with another about the outcome of a measurement.  Note that, since   it is arguably the case that standard quantum mechanics does not offer unequivocal answers about what will be witnessed by the various different observers in a Wigner's Friend or Extended Wigner's Friend scenario, we will here make use of the concept of an `extension' of quantum mechanics, by which we mean simply some  probabilistic or deterministic specification of what all of the observers will see in these scenarios. Since we are imagining that all MOPO approaches have been ruled out, we will require that an extension of quantum mechanics  specifies no more than one outcome per observer for each measurement. Thus we can specify our three types of `disaccord'\footnote{Note that for the sake of conciseness, in these definitions we use the phrase `does not agree' to mean either that two things can be compared and they are different, or that they just cannot be compared at all. This is why we have used the term `disaccord' rather than `disagreement'  -  `does not agree' is not necessarily synonymous with `disagree' here, since it also includes cases where there is neither agreement nor disagreement.}:

\begin{enumerate}

\item \textbf{Type-I Disaccord:} Suppose that Divya performs some measurement $D_1$ and then Alice performs some measurement $A_1$ and subsequently applies  quantum mechanics to make some inference about the result of Divya's measurement $D_1$. We will say that an extension of quantum mechanics exhibits Type-I disaccord if it tells us that Alice’s inference may not agree with what Divya herself experienced the outcome of measurement $D_1$ to be, even if Alice applied quantum mechanics correctly. 

\item \textbf{Type-II Disaccord:} Suppose there is some quantum state $\psi$ which gives the best possible predictions for all the results of all   the measurements that Alice could possibly perform on Chidi at some time $t$; and suppose this quantum state $\psi$ is naturally interpreted as representing something about Chidi's experiences at or shortly before the time $t$. We will say that an extension of quantum mechanics  exhibits Type-II disaccord if it tells us that the `natural' representation of Chidi's experiences suggested by the state $\psi$ may not agree with what Chidi is actually experiencing at or shortly before the time $t$. 
 
\item \textbf{Type-III Disaccord:} Suppose Chidi performs some measurement $C_1$, and then  Alice performs a measurement $A_2$ on Chidi himself, with the aim of establishing what his measurement result was - for example, this could involve simply asking him what the result was. We will say that an extension of quantum mechanics exhibits Type-III disaccord if it predicts that even under the most favourable conditions possible, the result of Alice’s measurement $A_2$ may not agree with what Chidi himself experienced the result  of the measurement $C_1$ to be.  
\end{enumerate} 

We will defer the discussion of Type-I disaccord to section \ref{OB}. For now, we will proceed as follows: first we will demonstrate that   the Wigner's friend scenario and the Bong et al and Lawrence et al theorems can be understood as demonstrating that any extension of quantum mechanics with certain properties (i.e. upholding the universality of unitary quantum mechanics,  and No-Superdeterinism, and  Locality,) must exhibit Type-II and/or Type-III  disaccord. We will then argue that, even if the Everett picture and other MOPO approaches are off the table, the existence of Type-II or Type-III  disaccord does not entail the existence of some kind of metaphysically radical non-absoluteness; there is always the option to adopt a  deflationary interpretation of disaccord which may have a relational flavour but which maintains that measurements have only one outcome for each observer \emph{and} that reality is made up out of objective, `absolute' facts.  Thus we will argue that these theorems do not in and of themselves provide any compelling scientific reason to believe in metaphysically radical non-absoluteness; that is a further interpretational choice which is not mandated by the scientific facts as they are currently understood.

\subsection{Type-II Disaccord\label{TypeII}}

The original Wigner's friend paradox   demonstrates that  any extension of quantum mechanics which is consistent with the universality of quantum mechanics in the first-person sense, but which also maintains that 
 every observer always sees a single definite outcome to any measurement they perform, must exhibit Type-II disaccord. This follows from the fact that the universality of quantum mechanics in the first-person sense entails that Alice's measurements on Chidi in the Wigner's Friend scenario must exhibit the statistics we would expect to see if Chidi were in the state $\psi$. Of course, in and of itself this statement about Alice's measurement results is nothing more than a description of  the dynamics of the Alice-Chidi interaction, which does not  entail any kind of disaccord:  if everyone in this scenario knows that unitary quantum mechanics is universal in the first-person sense, then everyone can agree that indeed, $\psi$ is a correct description of the dynamics of Chidi and his system with respect to all of the measurements that Alice is able to make. But we will potentially get disaccord if  Alice begins making additional interpretative assumptions about the meaning of the state $\psi$; for the state $\psi$  appears to have the form of a superposition of different conscious states, so if Alice observes statistics consistent with Chidi being in the state $\psi$, the natural assumption for her to make is that he has not in fact experienced any definite outcome.  And if she does assume this then she will end up making an inference about Chidi's experiences which does not agree with  what Chidi himself has actually experienced, since ex hypothesi he has in fact seen a single definite outcome - thus we end up with  Type-II disaccord.

Now, as  we have just described this scenario, there does not seem to be any metaphysically radical non-absoluteness involved - Alice is simply making an incorrect interpretative assumption which leads her to hold some incorrect beliefs about what Chidi has experienced. But some proponents of relational and perspectival interpretations would like to make a stronger claim: they argue that Chidi \emph{really} has not seen any definite outcome relative to Alice, although  of course he has seen a definite outcome from his own point of view, so there is just no fact of the matter about whether or not Chidi has seen a definite outcome. 

To analyse this claim, it will be helpful to be more specific about the meaning of the word  `state.' For this term is often used in  a way which equivocates between two possible meanings:  a) an intrinsic description of    a system at a given time or over some relatively short time interval, and b) a tool for predicting the outcomes of measurements performed on a system  at some time or over a relatively short time interval. For example, quantum states are often regarded as being both an ontological representation of a system at a time, and also a tool which encodes all available information about  the results of possible measurments on that system at that time.  In this paper we will refer to a) as the intrinsic condition, and b) as the dynamical state, because in a theory which incorporates measurements into its physical description rather than treating them as exogenous, a tool for predicting the outcomes of measurements on a system must ultimately be thought of as providing  a description of the  dynamics for the interaction of the system with some other system which can be understood as `measuring' it.

 Note that our term `intrinsic condition' is intentionally vague - it is supposed to refer to some objective, observer-independent description of the relevant system, but we are trying to avoid assumptions about the nature of that description. It might perhaps be like a classical state, but might also be something quite different - for example, in the Bell flash approach\cite{Gisin2013}, in which reality is composed of a distribution of pointlike events across spacetime, the `intrinsic condition' of a system at a time would simply be the distribution of flashes across some spacetime region which is roughly occupied by the system. We will, however, stipulate that for a system which  is a conscious observer, if this observer has anything which can be described as an intrinsic condition, then her intrinsic condition is what determines her conscious experiences and observations\footnote{We don't know what would determine the conscious experiences and observations of an observer in a radically perspectival picture in which there cannot be any intrinsic conditions; we leave that for the proponents of radically perspectival pictures to specify.}. This seems like a reasonable stipulation at least for those of a physicalist persuasion, since the nature of the `intrinsic condition' can be made quite general to accommodate various different theories of the physical basis of consciousness. 

Thus our starting point will be that for a conscious observer, if she has an intrinsic condition then her intrinsic condition determines her experiences, whereas her   dynamical state determines the statistics for the outcomes of measurements performed on her by other observers. But because it is common to refer to both the intrinsic condition and the dynamical state as simply `the state,' it is also common to assume that there is a straightforward mapping between the two such such  that the conscious experiences of an observer can simply be read off  her \emph{dynamical} state. Thus for example, in section \ref{Intro} we defined the set of states $ \{ | C_i \rangle \}$ by saying that `$|C_i \rangle $ \emph{is the state of Chidi corresponding to him having seen the outcome $|S_i \rangle$}.' Here the states $|C_i \rangle$ are being used as descriptions of Chidi's intrinsic condition, but we also stipulated that the state   $\psi = \sum_i c_i |C_i \rangle |S_i \rangle$ describes the measurement outcomes that will be obtained when Alice measures Chidi, and thus the states $| C_i \rangle$ are also being used to construct the dynamical state. This conflation of the intrinsic condition  with the dynamical state is entirely standard practice within the field of quantum foundations and beyond - and yet it must be kept in mind that this standard practice is based on a number of interpretative assumptions, as we are   not logically compelled to assume any particular link between the intrinsic condition and the dynamical state. 

Returning to the Wigner's friend scenario, then, the fact that Alice's measurements  exhibit the statistics we would expect for Chidi being in state $\psi$ can be expressed by saying that the dynamical state of Chidi  is $\psi$ (or at least, that is his dynamical state relative to Alice, though we should leave open the possibility that he has a different dynamical state relative to other observers). Then if it is assumed that Chidi's dynamical state must closely reflect his intrinsic condition, it is natural to conclude that Chidi's intrinsic condition  is some kind of indefinite superposition, so he has not made any definite observation. But at the same time, it has been stipulated that from his own point of view Chidi has indeed made a definite observation, and this is what gets  us to the supposed `non-absoluteness of observed events' advocated by proponents of relational and perspectival approaches.

However, the supposed inevitability of this non-absoluteness rests on equivocation between the intrinsic condition and the dynamical state. For we are not compelled to assume that Chidi's dynamical state is related in this way to his intrinsic condition; we could alternatively say that Chidi has a well-defined, observer-independent intrinsic condition according to which he has seen a definite outcome, but the intrinsic condition of Chidi does not map to the dynamics of the Alice-Chidi interaction in the way we would naturally expect, so despite the fact that Chidi has seen a definite outcome $|S_i \rangle$, his dynamical state relative to Alice does not take the form $|C_i \rangle$. Evidently there is nothing non-absolute about any of this, and nothing prevents us from giving an objective, third-person description of such a scenario - the  description will simply specify some definite observed measurement outcome for Chidi and then also note that the dynamics of the Alice-Chidi interaction are as described by the dynamical state $\psi$, despite the existence of that definite observed measurement outcome.

Of course, although the postulation of non-absoluteness is not compulsory here, one may still voluntarily choose to make interpretative assumptions about the existence of non-absoluteness. For example, one may postulate that the \emph{reason} the Alice-Chidi interaction is described by the state $\psi$ is because the intrinsic condition of Chidi, relative to Alice, is indefinite, even though from his own point of view he is in a single definite state. Or alternatively one may postulate that the reason the Alice-Chidi interaction is described by the state $\psi$ is because there is no such thing as an intrinsic condition, so there is nothing to be said about this scenario beyond the specification of the dynamical state.  But it must be emphasized that these kinds of assumptions  are not necessary to resolve the Wigners’s friend paradox, as we have already resolved the paradox simply by postulating that Chidi can have a dynamical state $\psi$ relative to Alice even though Chidi himself has seen a definite outcome. There is no need to say anything at all about Chidi's experiences or outcomes `relative to Alice' - perhaps the motivation for this is the idea that these somewhat unusual dynamics can't be explained without some kind of metaphysically radical relativization, but we do always have the option of simply  regarding all of this as a brute fact about the dynamics of the theory, or perhaps appealing to some axiomatization of quantum mechanics to explain why the dynamics are the way they are.

Moreover,  it’s not obvious what meaning  claims about Chidi being in an indefinite state `relative to Alice' are supposed to have. After all, in this kind of relational picture it is explicitly assumed that we are not working in a many-worlds scenario in which   different copies of Chidi can have different conscious experiences, so we know that Chidi can only have one set of experiences, which would seem to suggest that he has only one intrinsic condition which cannot really be relativized to anyone. If there are  `copies' of Chidi in some other kind of condition in versions of reality defined relative to other observers, and those copies are in some sense physically real, why are those copies not also having conscious experiences?  Similarly, it's not straightforward to make sense of the claim that reality   is composed entirely of dynamical states without any intrinsic conditions at all, for dynamical states are nothing more than sets of possibilities for future interactions, and it's certainly controversial to suggest that there could be a reality containing nothing other than possibilities. And it must be emphasized that all the relevant empirical content is  already contained in the assertion that the state of Chidi relative to Alice is $\psi$ - further statements about what Chidi has or has not seen relative to Alice tell us nothing about what either Chidi or Alice will experience.  So this statement about Chidi having a condition relative to Alice which is different from his actual experiences seems to float free from reality, describing nothing but experiences that nobody can possibly have, and thus in addition to being unnecessary it’s arguably incoherent. 

Possibly  the claim about Chidi's measurement outcome `relative to Alice' should really be understood as asserting that the inference Alice makes about Chidi's outcome is not \emph{just} a mistake in the ordinary sense. To make this point it is helpful to distinguish between two different kinds of mistake Alice could hypothetically make. First, she could think that the dynamical state of Chidi and $S$ relative to her is $\psi$ when really it is some basis state $| C_i \rangle \otimes | S_i \rangle$ - and if she makes this mistake,  some of her predictions for the results of measurements on Chidi  and $S$ will turn out to be wrong, so such mistakes have real empirical consequences. Clearly this is quite different from the kind of mistake where Alice thinks Chidi has not had a definite experience when in fact he has, since this belief will never cause Alice to make any predictions which could be empirically falsified. So the statement that Chidi has no definite measurement outcome `relative to Alice' could be understood as a kind of shorthand intended to express that Chidi's definite outcome is not dynamically relevant to Alice and thus in a sense she is not making a mistake when she says to herself that he has no definite measurement outcome, even though he has in fact experienced an outcome. If this is what is meant by the terminology, we have no particular objection - we would simply  caution that this way of using the phrase `relative to Alice' means that the  relativization of states  does not entail the existence of any kind of metaphysically radical non-absoluteness, since the whole story is entirely compatible with the idea that there is an objective, mind-independent fact about Chidi's outcome which just happens to be dynamically irrelevant for Alice. 

So where does the talk of radical non-absoluteness come from? It could be that there is a  mildly verificationist attitude at play here: if one believes that  what is true for Alice is defined by what she can measure, then indeed to all effects and purposes Chidi has `really’ not had a definite experience relative to her. However, surely a fully-fledged verificationism would lead us to question the wisdom of Alice holding any beliefs about Chidi's measurement outcome at all. After all, she   can get all the predictions she needs from the claim that Chidi has dynamical state $\psi$ relative to her, so further claims about his measurement outcome, his experiences, or his intrinsic condition are not useful for her predictions and indeed are not  really meaningful for her according to the verificationist criterion of meaning.   Thus even from this verificationist point of view, there’s no clear reason to think that Chidi having dynamical state $\psi$ relative to Alice must be interpreted in terms of Chidi being in some kind of indefinite condition `relative to Alice,' and thus there’s still no need to understand what is going here in terms of metaphysically radical non-absoluteness.

\subsection{Type-III}

The Bong et al experiment can be used to demonstrate that  any extension of quantum mechanics which is consistent with the universality of unitary quantum mechanics in the first-person sense,  which  maintains that 
 every observer always sees a single definite outcome to any measurement they perform, and which obeys the assumptions of Locality and No-Superdeterminism, must exhibit Type-III disaccord. The Lawrence et al theorem demonstrates roughly the same thing. We will focus on the Bong et al theorem in this section, but we think most of our conclusions would carry over to the Lawrence et al theorem.   

For our purposes, the important thing to notice about these two theorems is that their AOE assumption is really two entirely separate assumptions\footnote{In the Lawrence et al theorem the argument is less direct: what we have called AOE1 is understood to follow from the   `existence of relative facts,' and the assumption that we have called AOE2 is never referenced explicitly, but seems to be assumed in the authors' discussion of the version of relational quantum mechanics with cross-perspective links. In any case, as we saw in section \ref{Intro}, if the Lawrence et al theorem is to be repurposed as a `non-absoluteness' theorem in the tradition of the Bong et al result, it does  require both assumptions AOE1 and AOE2, since otherwise several of the trios of outcomes to which the quantum constraints are applied could never be accessible to any individual observer, and therefore the universality of quantum mechanics in the first-person sense would not imply anything about them.}. The first assumption, which we will henceforth refer to as AOE1,  is that  measurements have exactly one outcome relative to any given observer; the second assumption, which we will henceforth refer to as AOE2, is that if one observer performs a measurement on another aiming to establish the outcome of their  measurement, the value  obtained by the first observer will be the same as the value that the second observer actually witnessed\footnote{Ref \cite{schmid2023review} also distinguishes between these two assumptions, referring to the first as `Absoluteness of Observed Events' and the second as `Tracking.'}. AOE1 is used by Bong et al to rule out MOPO approaches, allowing them to insist that there are exactly four measurement outcomes witnessed by the people actually performing the measurements on each run of the experiment, and thus there must exist a well-defined probability distribution over the four measurement outcomes obtained by the people performing the measurements  in any run of the experiment. Meanwhile, AOE2 is what makes it possible to argue that Alice can access to Chidi's measurement and Bob can gain access to Divya's measurement outcomes and Alice and Bob can share measurement outcomes with one another,  and therefore the probability distribution over four measurement outcomes as obtained by the people who originally performed the measurements  must obey all the predictions of quantum mechanics for  any pair of values in the set $\{ AB, AD,   CB,   CD\}$. We emphasize the importance of AOE2 here - if it were not the case that observers are able to share their original measurement outcomes with each other, then nobody in this scenario would ever have access to more than one of the original measurement outcomes $A, B, C, D$ , and therefore nobody would ever need to see anything incompatible with quantum mechanics even if  the probability distribution over the values for the pairs $\{ AB, AD, CB, CD \}$ were not as predicted by quantum mechanics. Thus without AOE2, the universality of unitary quantum mechanics  in the first-person sense could not be used to used to arrive at any contradiction with the other assumptions.  Subsequently other authors have considered relaxing this assumption to arrive at inequalities applying to scenarios in which Alice can gain only incomplete information about Chidi's measurement outcome\cite{Moreno_2022} and similarly for the other observers, but of course in these cases it is still assumed that \emph{some} information can be obtained; no theorem of this kind can be proved without Alice having some sort of access to Chidi's observations and Bob's observations and to Divya's observations via Bob. 

Since AOE is really two separate assumptions, it follows that if experiments of the kind envisioned by Bong et al were in fact to  have the results predicted by unitary quantum mechanics, then in order to explain how this could come about we would only need to reject one of the two assumptions AOE1 abd AOE2. In particular,    since we are interested here in whether the non-absoluteness theorems can be used to argue for some kind of metaphysically radical non-absoluteness rather than simply a MOPO approach, we are interested in the case where AOE1 is retained and AOE2 is rejected; and evidently the rejection of AOE2 means that we have Type-III disaccord. Thus   one way of interpreting the Bong et al  theorem is to see it as an argument that if unitary quantum mechanics is universal in the first-person sense, and measurements always have a single outcome for each observer, and  the Locality and No-Superdeterminism assumptions are both correct, then  Type-III disaccord must exist.

It should be noted that the assumption AOE2 is also needed for a variety of other no-go theorems, most notably Bell's theorem - for if we can't assume that the two observers in a Bell experiment are able to  compare their results, we can't conclude anything about nonlocality.   Thus one might wonder why we are placing so much emphasis on AOE2 in the context of the Bong et al and Lawrence et al theorems, given that it is also needed by many other theorems. And in fact, we are focusing on AOE2 here  simply because denying AOE2 appears to be the route out of the Bong et al conundrum advocated by those who contend that the non-absoluteness theorems are compelling arguments for some kind of metaphysically radical relationalism or perspectivalism  - certainly, proponents of these views are not denying Locality, No-Superdeterminism, or AOE1, so it would appear that they \emph{must} be denying AOE2, and this would certainly fit with a radical perspectivalist approach in which facts defined relative to different agents may not be compared in any way. Thus, since our aim here is to understand whether  the non-absoluteness theorems really offer any compelling argument for these metaphysically radical possibilities, it is important for us to take seriously the possibility that AOE2 could fail and seek to determine whether that would really entail some kind of metaphysically radical picture. It is not, however, our intention to argue that denying AOE2 is a good or plausible idea  - indeed, shortly we will argue that it is neither. Our point in this section is simply to show that even if one \emph{does} go down the route of denying AOE2 and accepting Type-III disaccord, that doesn't inevitably lead to metaphysically radical relationalism or perspectivalism. 
 
Of course, there is no doubt that Type-III disaccord would be quite strange, but this does not mean it must be interpreted along metaphysically radical lines.  Bong et al seem to be  assuming, or perhaps defining, that an observed event is `absolute' only if information about it can be accessed by observers other than the person who initially observed it; and indeed, this idea was subsequently formalised  in ref \cite{Moreno_2022}, which defines a `non-absoluteness coefficient' quantifying the extent to which the outcome of Alice's measurement on Chidi is correlated with the result that Chidi himself obtained. But   the failure of `absoluteness' in this sense does not  entail that there is anything radically relational or perspectival going on - after all,  most  realists would presumably accept that there can  be a well-defined, objective fact of the matter about  what has been observed by some observer even if no other observer ever finds out about it, or could possibly find out about it. Moreover, the definition used in Bong et al arguably doesn't reflect  what most people intuitively imagine `absoluteness of observed events' to mean. Indeed, Bong et al themselves initially define AOE as the requirement `\emph{an observed event is a real single event, not relative to anything or anyone}'  and yet the denial of AOE2 and hence the existence of Type-III disaccord seems entirely compatible with this kind of absoluteness. For example, Ormrod and Barrett use the same preliminary statement of the meaning of AOE, but then cash it out as requiring that there exists  a fact of the matter about the measurement outcome which could in principle be written down as part of a non-relativized list of outcomes; and Type-III disaccord is clearly not incompatible with this, as the fact that Alice cannot find out Chidi's outcome does not entail that the  outcome could not in principle be written down in a non-relativized list of outcomes, although obviously \emph{Alice}  could not write it down. 

Indeed,  on closer examination it is clear that Type-III disaccord is really just a special case of Type-II disaccord, which of course need not involve metaphysically radical non-absoluteness. The only real novelty  here  relative to the Wigner's Friend case is that we are now allowing  the set of possible Alice-Chidi interactions to  include an interaction which is naturally described as `Alice asking Chidi the outcome of his measurement.' So we can simply respond by saying, as in the Wigner's friend case, that in this experiment the dynamical state does not map onto the intrinsic condition in the way one would ordinarily expect, and in particular the dynamical state of Chidi relative to Alice is $\psi$  even though Chidi's intrinsic condition is that he has obtained a definite outcome to his measurement. This immediately entails the the existence of Type-III disaccord, for clearly if Chidi's dynamical state relative to Alice is $\psi$ then the specific value of his outcome is not dynamically relevant for her, and therefore no possible interaction with Chidi can provide Alice with reliable information about the value of his outcome, even though there still exists an objective fact of the matter about what that outcome was.   From this point of view, the problem is not that Chidi's outcome is not `absolute'  - the difficulty is simply that the dynamics are defined in such a way that Alice is not able to access that absolute outcome through any physical interaction, even one as straightforward as having a conversation with Chidi.  So relativizing the \emph{dynamical} states gives us everything we need to make sense of the Bong et al result, without introducing any kind of metaphysically radical non-absoluteness. 
 
Of course, perhaps those who advocate metaphysically radical non-absoluteness have in mind  some kind of verificationism which has the consequence  that if observers can’t possibly compare their measurement outcomes, then it isn’t meaningful to talk about the relation between their measurement outcomes at all, which would seem to justify the claim that there can’t be any absolute description of the outcomes. But clearly this approach is not \emph{mandated} by the existence of Type-III disaccord as manifested in the Bong et al experiment, so the theorem by itself provides  little reason to take that route unless one already has verificationist sympathies.  Thus, although we do think that Type-III disaccord involves a more significant revision of our intuitive picture of the world than Type-II or Type-I, nonetheless  we are still not convinced that the existence of Type-III disaccord provides any compelling reason to postulate the existence of metaphysically radical non-absoluteness.

\section{Intersubjectivity}

We should now address a potentially serious problem for approaches in which dynamical states fail to map onto intrinsic conditions in the way one would naturally expect. That is, one might worry that allowing  a mismatch between dynamical states and intrinsic conditions will lead to a kind of extreme solipsism - for after all, the dynamical state of an observer determines the outcomes of your interactions with her, so if her dynamical state is completely unrelated to her intrinsic condition, there is no possible way for you to ever get to know anything about her intrinsic condition. This would lead to a picture of reality in which each one of us is trapped inside of our own separate reality, with no ability whatsoever to learn anything about what is going on inside someone else’s reality -  which is exactly the  problem envisaged in ref \cite{https://doi.org/10.48550/arxiv.2203.16278}, where it is argued that such an extreme failure of intersubjectivity would undermine the practice of science. So it seems that allowing a mismatch between dynamical states and intrinsic conditions could in the end be just as bad as metaphysically radical non-absoluteness in terms of its consequences for the epistemology of science. 

Thus in order to arrive at a viable account of the Wigner's friend and extended Wigner's friend scenarios by means of separating out dynamical states from intrinsic conditions, we probably need to maintain that in general there exists some systematic relation between the dynamical state of a system and the intrinsic condition of that system. But fortunately, that leaves a lot of freedom to fix the exact nature of this systematic relation.  In particular, for the purposes of upholding the rationality of science it  doesn't really matter whether or not `supermeasurements' yield reliable information about intrinsic states, given that no supermeasurement has ever yet been performed on or by any human being.   What matters for  scientific rationality is simply that ordinary kinds of physical interactions - the ones that real people have historically used and continue to use today  to establish a shared intersubjective reality - should in general deliver meaningful information about the intrinsic condition of other observers. 

That is to say, we can straightforwardly accommodate Type-I and Type-II disaccord in a scientific worldview, as long as we don't also have \emph{Type-III} disaccord, for it is only Type-III disaccord that would seriously undermine the reliability of the ordinary physical interactions that we use to establish our shared intersubjective reality.  For example, we have noted that in the Bong et al scenario Alice’s `measurement’ $A_2$ could simply take the form of asking Chidi what result he saw, so approaches allowing Type-III disaccord in this context are  committed to the claim that even when Chidi verbally tells Alice his result, the words that Alice hears may not be the same as the words that Chidi imagines himself to have spoken. Moreover, note that there is nothing special about this interaction - Alice is no longer maintaining coherent control over Chidi, so this is just an ordinary conversation between two people, in physical terms no different to any other conversation - and thus if we accept the possibility of Type-III disaccord in this context, we would apparently be forced to accept that quite generically, what you hear when I communicate with you may not be the same as what you imagine yourself to have said. 

So it is quite important to establish whether we can maintain the universality of unitary quantum mechanics in such a way that  dynamical states come apart from intrinsic conditions just enough to give us Type-I and/or Type-II disaccord, but not Type-III disaccord. And in fact the Bong et al theorem answers this question: if we are not willing to accept a MOPO approach,  and we are also determined to maintain Locality and No-Superdeterminism, it is impossible to achieve such a thing. 

Now, one might be tempted at this juncture to think that this is a conclusive argument against the approach we have suggested for maintaining absoluteness in the face of the Bong et al theorem, and thus the natural response is to go back to non-absoluteness. But these difficulties are not solved by postulating some kind of metaphysically radical absoluteness. Quite the reverse: once we deny the possibility of third-person, objective descriptions of reality, Type-III disaccord will be \emph{everywhere}, because in such a context  there is no hope whatsoever that any observer can reliably find out about anything which has been witnessed by other observers. The existence of reliable links between perspectives requires the existence of facts about the relations between perspectives which are not themselves relativized to anything else; in order to maintain intersubjectivity we do need some bedrock of objective, non-relative facts, even if those facts are entirely facts about relations (see ref \cite{https://doi.org/10.48550/arxiv.2203.16278} for a detailed account of the problems for the epistemology of science which arise from denying the existence of any non-relative facts). So the really essential point here is that there is nothing to be gained   by adopting metaphysically radical non-absoluteness as a response to the Bong et al theorem: we get Type-III disaccord either way. And in our view, accepting the existence of Type-III disaccord is simply not  a viable option, since the violence that it would do to the epistemology of science is beyond what can reasonably be tolerated. 

Thus we conclude that the right response to the Bong et al theorem cannot be to reject AOE2. Moreover, we have stipulated that for the purpose of the current disussion MOPO approaches have been ruled out, which means that we cannot reject AOE1 either. So we cannot respond to the Bong et al theorem by rejecting either part of AOE  - if unitary quanutum mechanics is in fact universal, one of the other assumptions must be rejected instead.  So assuming that we are still insisting on the universality of unitary quantum mechanics in the first person sense,  it seems the only path forward  would involve rejecting either Locality or No-Superdeterminism - which, recall, is written by Bong et al in such a way as to rule out both superdeterminism and retrocausality. Therefore in our view the Bong et al theorem does not really offer any argument in favour of metaphysically radical absoluteness - it is much more naturally construed as an argument in favour of the failure of Locality, superdeterminism or retrocausality! 

That said, of course the option remains open here to adopt the Everett approach or some other MOPO approach rather than accepting superdeterminism or retrocausality, and we suspect this  will be many people's first choice. But we would caution against too quickly retreating to a MOPO approach. In particular, we want to emphasize that the  `retrocausality' that would be needed to avoid Type-III disaccord in the Bong et al and Lawrence et al experiments is of a very special kind, and some of the usual reasons one might have for wishing to rule out retrocausality do not apply in this context. Indeed, we think  the fact that the `retrocausality' needed to avoid Type-III disaccord in the Bong et al experiment and Lawrence et al experiments happens to be of this very special kind is not just a coincidence - rather it should be taken as an indication that  `retrocausality' may be precisely the right solution to the problem posed by these theorems! So let us now consider some obvious objections to retrocausality and show that they do not apply to this special kind of `retrocausality.'

\subsubsection{Paradoxes \label{retro}}

One of the main reasons for being wary of retrocausality is the fact that it can in principle be used to create logical contradictions. For example, it is easy to construct a retrocausal version of the grandfather paradox: suppose that instead of time traveling, our intrepid experimenter makes use of a retrocausal mechanism to cause her grandfather to die before he can father any children. But then the time traveler will not exist, so she won’t be able to cause her grandfather’s death.  Thus if the retrocausal mechanism is deterministic and perfectly reliable, there is no logically consistent way to resolve this set of events; so evidently this kind of retrocausal mechanism cannot exist if the world cannot contain logical contradictions. 

Of course, many of the processes we are concerned with in quantum mechanics appear to be indeterministic, so this kind of argument  may not directly apply. But we can easily imagine an indeterministic analogue, in which a set of probabilistic processes are composed in a way which makes it impossible for them all to exhibit relative frequencies exactly equal to or close to the values of their theoretical probabilities; we will refer to this as a `probabilistic contradiction.' For example, suppose we gather a collection of time travelers and have them make use of a retrocausal mechanism which is supposed to have fifty percent chance of causing the death of the person on whom it is directed at a chosen time. If all of the time travelers direct this mechanism on their grandfathers at a time before these grandfathers have fathered any children, we would naively expect around fifty percent of the grandfathers to die. But in fact, it is not logically possible for any of the grandfathers to die, because any grandfather who dies before he fathers any children cannot have a grandchild who uses this retrocausal mechanism on him. So no matter how many times we try this experiment, the relative frequency of death will always be zero, which is very far from the expected frequency of fifty percent. Now, probabilistic contradictions are evidently not impossible in the way that logical contradictions are, but there is  still something undesirable about them: for surely any reasonable account of probability would tell us that if a probabilistic process reliably and robustly deviates from its theoretical probabilities in a certain  context, then those theoretical probabilities are simply not the right ones, and   we are really just dealing with a different process altogether. Thus it seems reasonable to think that retrocausal mechanisms also cannot exist if they could be used to create probabilistic contradictions, since they will necessarily just become  different mechanisms which don't produce contradictions. 

Now, observe that the kind of retrocausality or superdeterminism we need to avoid Type-III disaccord in the Bong et al scenario is simply any kind of  relation such that Chidi's measurement outcome is not independent of  Alice's choice of measurement (and/or likewise mutatis mutandis for Divya and Bob). This relation need not necessarily be strictly `causal,' but it is clearly a dependence relation going backwards in time, and in that sense it might be described as retrocausal. But note that this particular instantiation of  backwards-in-time dependence  cannot possibly be used to create logical contradictions or probabilistic contradictions! For example, to create a logical contradiction using the dependence of Chidi's measurement outcome on her choice of measurement, Alice would have to note that some particular result for Chidi’s measurement is supposed to occur if and only if she performs some particular measurement (WLOG say Chidi’s measurement has the result $+1$ if and only if Alice performs $B_1$), and subsequently  perform an experiment where she first learns the result of Chidi’s measurement and then, if the result is $+1$, she deliberately chooses to perform measurement $B_2$, meaning that the result of Chidi’s measurement must be $-1$ even though Alice already knows it is $+1$. Similarly, to create a probabilistic contradiction using the dependence of Chidi's measurement outcome on her choice of measurement, Alice would have to note that some particular result for Chidi’s measurement is supposed to occur with probability significantly less than $1$ whenever she performs some particular measurement (WLOG say Chidi gets result $+1$  with probability $<<1$ if Alice performs measurement $B_2$), and subsequently perform a series of experiments where she first learns the result of Chidi’s measurement and then, if and only if the result is $+1$, she then performs measurement $B_2$, so Chidi always gets result $+1$ when Alice performs measurement $B_2$, even though the probability of this occurrence is supposed to be significantly less than $1$.  

The key point here is that for either a logical or probabilistic contradiction, Alice must first learn the result of Chidi’s experiment and then decide on this basis which measurement to perform. This is precisely analogous to the grandfather paradox, or its retrocausal version – in effect both paradoxes begin with the experimenter learning whether or not her grandfather died before fathering children, since she knows he cannot have done so if she herself exists. And then clearly she subsequently kills her grandfather if and only if he didn’t die before fathering children, since she can’t exist or kill anyone if he did die before fathering children. So as a general rule, the construction of this kind of contradiction requires observers to be able to find out about the past events featuring in the contradiction, in order that they can carry out actions which contradict the events they have learend out. 

But quantum mechanics tells us that Alice can perform a `supermeasurement’ on Chidi only if she preserves him in a coherent state inside his laboratory. And preserving him in a coherent state means ensuring that no information from inside of his laboratory escapes to the outside, to be learned by Alice or anyone else. So Alice cannot first find out the result of Chidi's measurement and then decide to perform either measurement $A_1$ or $A_2$ - for learning the result of Chidi's measurement \emph{just is} performing measurement $A_2$, and once measurement $A_2$ has been performed Alice cannot also choose to perform the supermeasurement $A_1$, as she must give up coherent control over Chidi in order to learn what his measurement outcome is. And therefore if there were a `retrocausal' dependence of Chidi's outcome on Alice's choice of measurement, in violation of the `No-Superdeterminism' assumption, Alice would never be able to use this to create logical or probabilistic paradoxes\footnote{A similar point was made by Price in ref \cite{Price_1994} about the kind of retrocausality needed to preserve locality in Bell scenarios, but we are not limiting ourselves to local retrocausality here}. 

So why should we be worried about this kind of retrocausality? There is a certain kind of temporal prejudice which inclines people to think that `retrocausality’ is just impossible – it is not the kind of thing the universe could possibly contain. But here is an alternative proposal: retrocausality is impossible \emph{precisely when it could be used to create logical and/or probabilistic  contradictions}. This proposal  implies that retrocausality \emph{should} be possible in special circumstances where it cannot be used to create logical or probabilistic contradictions - and that  is exactly what we see if we try to introduce retrocausality in order to  avoid Type-III disaccord in the Bong et al scenario, so from a certain point of view it would not be so shocking to discover that some kind of retrocausal effect occurs in this setting.

\subsubsection{Differing conceptions of retrocausality \label{retro2}}

Another reason for being wary of retrocausality is that the very idea of it seems suspect in a metaphysical or ontological sense. Are we really supposed to posit two different, opposing arrows of causality which somehow `collide’ in the middle to determine intermediate measurement outcomes? To many people this sounds ludicrous. Furthermore, many philosophers take the view that `causation’ is really a macroscopic phenomenon associated with the temporal direction derived from the thermodynamic gradient\cite{Price2005-PRICP}, and to someone who understands causation in this way, it makes no sense to imagine that causation could proceed backwards in time from Alice's choice to Chidi's outcome: causation is tied to the thermodynamic gradient, and the experiments involved in the non-absoluteness theorems do not involve any abnormal fluctuations of entropy which could conceivably be seen as reversing the thermodynamic gradient.

However, the kind of `retrocausality’ needed to resolve the non-absoluteness theorems need not be understood in terms of a literal backwards arrow of causality. All we need is to say that Chidi’s result may depend on Alice’s choice - it is not necessary to interpret this dependence as  causal  in any strong sense. For example, ref \cite{adlam2022roads} argues that the right way of understanding this kind of dependence is to think of the entire set of events as being determined `all-at-once,’ so there is neither a forwards nor a backwards arrow of causality at the fundamental level; rather the past and the future mutually depend on one another, in which case we would naturally expect to find instances where there is something like a dependence relation going backwards in time. In such a picture it is no trouble at all to have Chidi’s outcome depending on Alice’s choice, since the events are mutually adjusted to fit the constraints imposed by the relevant quantum-mechanical laws, which we now know will always be possible since we have just seen that the dependence of Chidi's outcome on Alice's choice cannot possibly produce any contradictions. And note that this fundamentally symmetric picture is very friendly to views of causation which see it as a macroscopic phenomenon associated with the thermodynamic gradient, as discussed in greater detail in refs \cite{Adlam2023-ADLITC-3}. So metaphysical or ontological objections to the notion of retrocausality are not necessarily good reasons to rule out a dependence of Chidi's outcome on Alice's measurement in the Bong et al scenario.

\section{Relational dynamic states}

Based on the preceding discussion, we do not believe that the original Wigner's Friend scenario  or the more recent non-absoluteness theorems offer any compelling reason to postulate some kind of metaphysically radical non-absoluteness, even if one is determined to maintain the universality of unitary quantum mechanics and one has already ruled out all MOPO approaches. However, our discussion has nonetheless  shown that these theorems may suggest the existence of a less radical kind of non-absoluteness, according to which \emph{dynamical states} are relational whilst  intrinsic conditions and hence observed events are still objective and observer-independent. After all, even in a completely mundane world composed entirely of observer-independent, `absolute' events, we can  imagine a   scenario in which the predicted statistics for Bob's measurements on Alice are different from the predicted statistics for Chidi's measurements on Alice, meaning that Alice has different dynamical states relative to Bob and Chidi even though she still has just one intrinsic condition. And indeed, one way of interpreting the non-absoluteness theorems is to think of them as telling us that  if  quantum mechanics is universal in the first-person sense, it is inevitable that systems will sometimes have different different dynamical states relative to different observers. 

This possible avenue for making sense of the non-absoluteness theorems has probably been obscured by the long-standing tradition of conflating intrinsic conditions with dynamical states. And of course, we should acknowledge that historically there has been a good reason for treating the two as interchangeable - locality. For locality tells us that the interaction between a system and a measuring device cannot depend on anything other than the intrinsic conditions of the system and the measuring device and their relative arrangement, and therefore given the intrinsic condition of a system we can always write down a unique `dynamical state' which specifies  the outcome or probability distribution over outcomes that would be obtained for every local interaction this system could possibly have with a measuring device. Thus in a local theory, it will always be possible to construct a simple  map from the intrinsic condition of a system to its dynamical state in a way which is valid for any possible measuring instrument, so there is little harm in using just one mathematical object to represent both of them.

But in a non-local theory, such as quantum mechanics arguably appears to be, it need not be the case that the interaction between a system and a measuring device depends only on the intrinsic conditions of the system and the measuring device and their relative arrangement, so this simple prescription for mapping an intrinsic condition to a dynamical state may no longer work. And indeed, the Wigner's friend scenario and non-absoluteness theorems appear to be telling us that if unitary quantum mechanics is universal in a first-person sense this must be the case.   For example, in the Wigner's friend scenario after Chidi has measured system $S$, if unitary quantum mechanics is correct in the first-person sense, then the result that Chidi would get if he were to perform another measurement on $S$ may be different from the result that Alice would get if she performed the same measurement on $S$, even if the two of them were to use identically configured measuring instruments\footnote{Probably they could not actually use completely identical  measuring instruments, since Alice must perform her experiment in a way which maintains full coherent control over the Chidi-$S$ system, but perhaps she could arrange that at least the part of her device which interacts locally with $S$ is configured in the same way as Chidi's instrument would have been.}. This suggests that in a quantum context we will not in general be able to write down a unique `dynamical state' for a system specifying the probability distribution over  outcomes for every possible configuration of the measuring device, because we also need to take into account facts about the person who is operating or who will later read the measuring device, and/or facts about recent history, and/or facts about the broader context of the measurement, all of  which may influence the measurement interaction in a subtle spatially and/or temporally non-local way. So we will end up with a complex, non-local, and probably somewhat `retrocausal' dynamics for the theory - but this complex dynamics can still be understood as producing well-defined, non-relative events and states of affairs, so although the dynamics are unusual, the metaphysics need not be particularly radical. 

The special features of the dynamical state in the quantum context also help show why it is a mistake to elide the dynamical state with the intrinsic condition.  For intrinsic conditions (if they exist) are facts about individual systems in and of themselves, whereas dynamical states are facts about the relation between a system and a measuring device, an observation which  becomes less trivial in the non-local quantum context. This explains straightforwardly why dynamical states can be relativized even if intrinsic conditions are non-relative: a  measurement is a   dynamical interaction between the system being measured and the system doing the measuring,  so it is entirely reasonable that the outcome of Alice's measurements on Chidi may not be entirely determined by facts about Chidi, since features of Alice and her history and her broader context may also be relevant. Moreover, intrinsic conditions (if they exist) are features of a system at a time or at least over a relatively short temporal interval, whereas we know that in quantum mechanics dynamical states can only be established by performing tomographic measurements on a large number of identically prepared systems, so possibly they should perhaps be thought of in terms of some kind of spatially and temporally non-local coordination, rather than as features of individual systems. Thus although classical physics allowed us to get away with identifying the dynamical state with the intrinsic condition, in a quantum context it appears that intrinsic conditions and dynamical states are not at all the same kind of object - indeed, arguably a lot of misunderstanding may have arisen from the decision to call the  quantum wavefunction a `state' even though it is really a \emph{dynamical} state and therefore disanalogous in many ways from the traditional classical concept of state, which is closer to our notion of intrinsic condition.   

  Thus  we can certainly have a coherent and interesting `relational' view which is compatible with the universality of quantum mechanics in the first-person sense but which does not require any metaphysically radical non-absoluteness. For the hypothesis that the outcomes of measurement interactions are determined by the features, histories and/or broader context of both systems involved  is perfectly compatible with, and indeed \emph{requires}, the hypothesis that  these systems have well-defined non-relative features, histories and/or contexts which  determine the outcome of the measurement interaction. So metaphysically radical hypotheses  are   not necessary to maintain the universality of quantum mechanics in the face of the Wigner's Friend paradox: the suggestion that \emph{dynamical} states are relativized is already a powerful and far-reaching idea which can potentially resolve a number of puzzles about quantum mechanics without any need to deny the absoluteness of observed events.

\subsection{Universality versus completeness}

At this juncture one might object that postulating  intrinsic conditions for quantum systems which are distinct from the dynamical state (i.e. the quantum state) looks like it might be incompatible with the assumption of the universality of unitary quantum mechanics, which was a key motivation for   this dynamically relational approach in the first place.  However, it is important to distinguish between the proposition that unitary quantum mechanics is `universal' and the proposition that it is `complete.' To say that unitary quantum mechanics is `universal' in the first-person sense  is to make an assertion about the dynamics of the theory - i.e. that the unitary dynamics will always predict the right  measurement outcomes for any individual observer.  Whereas to say that quantum mechanics is `complete' is to say that once we have specified the quantum state of a given system (at a given time), there is nothing further to say about it (at that time) - and this is evidently a claim about kinematics or perhaps  ontology, rather than about dynamics. Furthermore,  we have seen throughout this paper that upholding the `universality' of quantum mechanics tends to lead towards views in which the quantum state is relational, whereas  once we start making quantum states relational  the claim that quantum mechanics is `complete' doesn't even seem well-formulated, because in that case one cannot simply specify the quantum state of a given system; one must always specify the state relative to some observer, and then any time we have specified a quantum state of a system there will always be  something further to be said about it, since it may have different states relative to other observers. So it seems clear that quantum mechanics can be `universal' in the sense that it always predicts the correct  measurement outcomes for any individual observer, even if it is not  `complete.'

Moreover, the scientific reasons for thinking that quantum mechanics must be universal, as discussed in section \ref{resp}, are principally to do with the problems that come from trying to stitch together two separate kinds of dynamics (e.g. the fact that we have no empirical evidence that the unitary dynamics of quantum mechanics ever gives way to any other dynamics, and the fact that it is difficult to combine two different dynamical processes together in a way that works for quantum field theory). So these reasons for thinking that quantum mechanics may be universal do not necessarily give us any reason to think that it must be \emph{complete}, for the claim that quantum mechanics is complete is an assertion about kinematics or ontology, and therefore denying it does not  entail that there must exist some second kind of dynamical process that we must integrate with the quantum dynamics. Thus there is arguably less justification for insisting on completeness than there is for insisting on universality in this context,  so we should  at least be open to  views where dynamical states diverge from intrinsic conditions in such a way that quantum mechanics is universal but not complete.

\section{Examples}

We have  offered a sketch of a solution to the measurement problem which, in virtue of making use of relativized dynamical states and a subtle kind of   retrocausality, is able to maintain the universality of unitary quantum mechanics and the existence of only one outcome per observer without postulating any metaphysically radical non-absoluteness. Lest anyone suspect that this is a mere fantasy, we will now give some examples of existing interpretations of quantum mechanics which implement this vision. 

\subsection{Relational quantum mechanics with cross-perspective links \label{RQM}}

Relational quantum mechanics (RQM), in its original form set out in ref \cite{dibiagio2021relational}, exhibits Type-III disaccord - for one of the postulates of RQM given in ref \cite{dibiagio2021relational} specifically tells us that it is not even meaningful to compare the perspectives of different observers, except by invoking a third system with respect to which the comparison is made. And clearly if we cannot even compare the perspectives of different observers from a non-relativized perspective, we cannot maintain that there is some physical interaction which reliably brings these perspectives into agreement from a non-relativized perspective, so Type-III disaccord must be generic in this picture. 

Recognising the epistemic problems posed by this failure of intersubjectivity, ref \cite{https://doi.org/10.48550/arxiv.2203.13342} proposes an alteration to RQM. The prohibition on comparing perspectives is removed, and instead a new postulate known as Cross-Perspective Links (CPL) is added: `\emph{In a scenario where some observer Chidi measures a variable V of a system S, then provided that Chidi does not undergo any interactions which destroy the information about V stored in Chidi’s physical variables, if Alice subsequently measures the physical variable representing Chidi’s information about the variable V, then Alice’s measurement result will match Chidi’s measurement result.}’ Ref \cite{https://doi.org/10.48550/arxiv.2203.13342} specifies that the information about V is destroyed (relative to Alice) precisely when Alice performs a measurement on Chidi which does not commute with the variable associated with that information – which is to say, once Alice has performed a supermeasurement, the information about Chidi’s measurement is no longer accessible to her in subsequent measurements.

 RQM+CPL exhibits Type-II disaccord: it tells us that in a Wigner's friend scenario, Chidi's intrinsic condition is that he has seen a single definite measurement outcome, but nonetheless   the dynamical state of Chidi relative to Alice is  the superposition state $\psi$, since over many repetitions of the experiment  Alice will see measurement outcomes consistent with the state $\psi$. Or rather, the dynamical state of Chidi relative to Alice is $\psi$  with one special exception   - if Alice measures in the basis corresponding to the variable of Chidi that records his measurement outcome, then she is guaranteed to get the same outcome as Chidi, even though a prediction based purely on the state $\psi$ would predict that other outcomes are possible. One might worry that this deviation from quantum mechanics will lead Alice to see outcomes which are incompatible with quantum mechanics, thus violating the universality of quantum mechanics in the first person sense, but this is not the case - for Chidi's outcome will be different on different runs of the experiment, and it follows from the linearity of unitary quantum mechanics that the expected statistics for Chidi's measurement are the same as the expected statistics for Alice's measurement in the corresponding basis on the state $\psi$, so over many repetitions of this scenario Alice will indeed see the outcomes expected from the state $\psi$\footnote{We have not been able to find any cases in which the cross-perspective-links postulate definitely leads to a prediction which is inconsistent with the predictions of unitary quantum mechanics, but no one has given a general proof of consistency, so it is possible that some inconsistency could still be found - it would certainly be a worthwhile project to investigate this further! However, we caution that attempts to show inconsistency should keep in mind that RQM+CPL can be formulated in a way that allows a subtle kind of retrocausality, which may help it avoid possible inconsistencies - indeed, we will shortly see an example of this with reference to  the Lawrence et al theorem.}. This goes back to the observation that dynamical states in quantum mechanics must be established using measurements on a large number of identically prepared systems - so perhaps we should say the state of Chidi relative to Alice really is just $\psi$ despite the exception we have just described, since what really defines the dynamical state is the ensemble statistics rather than the individual case. 
 
But in any case, by construction RQM+CPL does \emph{not} exhibit Type-III disaccord: CPL is specifically designed to  ensure that, when   Alice correctly measures the variable encoding Chidi’s measurement result, Alice’s measurement result will match what Chidi himself saw. The key point here is that although in RQM+CPL observed outcomes are dynamically relevant only to measurements performed in one special basis,  macroscopic observers almost exclusively end up measuring one another  in that special basis, because to use any other basis they would have to maintain complete coherent control in order to perform a supermeasurement, and decoherence makes this more or less impossible. Thus although the complete dynamical state of a  macroscopic observer relative to another will not usually be the eigenstate that we would naturally associate with  their intrinsic condition, nonetheless the full   dynamical state will generally be inaccessible to other observers because the tomographic measurements that would reveal it are almost impossible to actually perform, and thus we can expect to end up with a stable quasi-classical intersubjective reality despite the fact that dynamical states do not typically map onto intrinsic conditions in the way one would naturally expect. 

Note that since RQM+CPL does not exhibit Type-III disaccord and does not allow more than one outcome per observer, but nonetheless also maintains the universality of unitary quantum mechanics in the first-person sense, it must violate one of the other assumptions of the theorem of Bong et al. And in fact we think a sensible formulation of RQM+CPL would certainly violate No-Superdeterminism\footnote{It is also possible that it would violate Locality. Certainly, RQM+CPL is in a sense non-local, if it is understood in terms of an `all-at-once' model, since such a model allows generic correlations between spacelike, timlike and lightlike separated events without any need for local mediation. The question then is whether RQM+CPL violates Parameter Independence (PI), in which case it violates the Locality condition of Bong et al, or just Outcome Independence (OI), in which case it does not violate the Locality condition. As a matter of fact, we suspect that the distinction between PI and OI only makes sense in a causal, time-evolution model, so in the all-at-once context it may not be possible to separate the generic nonlocality out into PI and/or OI; thus it is likely that RQM+CPL would violate PI in a formal way, though in the all-at-once context that would not have the same implications it is claimed to have in the time-evolution context, and in particular it would not necessarily entail superluminal action or a preferred reference frame.}. Specifically,  the kind of subtle retrocausality discussed in section \ref{retro2} actually seems quite natural  in the context of RQM+CPL, because in that picture we are required to postulate the existence of a large network of \emph{absolute, observer-independent} events distributed over spacetime in such a way that they exhibit non-local quantum correlations across both spacelike and timelike  separations, and it is has been recognised in the context of the Bell flash ontology that in order to have relativistic covariance in a picture based on an ontology of correlated pointlike events, we typially have to `\emph{renounce any account of the coming-into-being of the (events)}.' That is, we must allow that the network of events is determined in an `all-at-once' manner, rather than being generated in some temporal order. And it was already noted in section \ref{retro2} that backwards influences which look superficially like retrocausality will likely be generic in an `all-at-once' picture, so we probably shouldn't be too surprised to learn that in RQM+CPL we will need to violate the No-Superdeterminism assumption.

This response also applies to the argument of Lawrence et al, which was originally intended specifically as a criticism of  RQM. As pointed out by ref \cite{cavalcanti2023consistency}, the argument of Lawrence et al clearly fails to land if it is directed at the original version of RQM, because this version of RQM tells us that the measurement outcomes obtained separately by Alice and Bob can never be compared, so the universality of unitary quantum mechanics in the first-person sense does not entail that these outcomes must obey the quantum constraints obtained by Lawrence et al. However, one might initially think that the argument would work if it is directed at RQM+CPL, since the whole point of adding the CPL postulate is to  allow observers to gain access to the observations of other observers. And indeed, Lawrence et al argue that in RQM+CPL all of the outcomes obtained by Alice and Bob are real for both of them: `\emph{in fact all outcomes exist for all observers, but some of them are just unknown}’ - and thus, they appear to think, in this case the six outcomes must definitely all of obey the constraints obtained from quantum mechanics. 

As a matter of fact, we do think it is correct to say that in RQM+CPL all outcomes exist for all observers, with some of them simply being unknown; and we also agree that in RQM+CPL, if   Bob measures Alice in an appropriate basis he can learn the values of $A_1, A_2$ and $A_3$, so  he could in principle access any one of the four trios in the set $ \{ B_1 B_2 B_3; B_1 A_2 A_3; A_1 B_2 A_3, A_1 A_2 B_3\}$. Thus the constraints derived by Lawrence et al will have to be satisfied for the trio that he does in fact access. However, Lawrence et al are wrong to assume that all four constraints must always be obeyed. Bob cannot access all four trios \emph{on any given run of the experiment}, since measuring $B_1$ is incompatible with measuring $A_1$ and so on. And in order to uphold the universality of unitary quantum mechanics in the first-person sense, RQM+CPL does not need to respect the quantum constraints for all outcomes which could in principle have been accessed by a given observer; it only has to respect the quantum constraints for all outcomes which  are \emph{actually} accessed by a given observer on a given run of the experiment. Thus even with CPL, universality in the first-person sense does not entail that all four trios must obtain the quantum constraints obtained by Lawrence et al, so the proof does not go through. 

Lawrence et al might try to argue, like Bong et al, that since Bob could measure any combination of these variables, then if Bob’s results always match the predictions of quantum mechanics we must be able to pre-specify values for all  six variables which obey the four constraints obtained from quantum mechanics, which their proof shows to be impossible. However,  as in Bong et al, we only need to be able to pre-specify values matching the quantum constraints for all four trios if we assume that the outcomes obtained by Alice are  independent of Bob’s subsequent choices and that Bob's individual outcomes $B_1, B_2, B_3$ are all independent of what other measurements Bob chooses to make, and thus this argument, like the Bong et al one, assumes Locality and No-Superdeterminism. Therefore since RQM+CPL is perfectly compatible with a subtle retrocausal effect of Bob’s choices on Alice’s measurements, it does not run into a contradiction here; so, contrary to their claims to have ruled out relative facts, the argument of Lawrence et al does not rule out either the original version of RQM or RQM+CPL, properly understood.

\subsection{Kent’s Lorentzian solution to the quantum reality problem}

Another way of implementing Type-II disaccord without Type-III disaccord involves postulating a model in which there are fewer observed events then we would naturally imagine. That is, we could say that there is not always an actual observed event associated with an instance in which the unitarily evolving wavefunction appears to represent an observer as making an observation, thus allowing us to evade the non-absoluteness theorems by saying that all observed events are absolute but some of the events featuring in these theorems are not actually observed. 

For example, this occurs in Kent’s Lorentzian solution to the quantum reality problem\cite{Kent,2015KentL, 2017Kent}.  Kent's proposed interpretation of quantum mechanics is based on a simple idea: there is no collapse of the wavefunction, so we just allow the wavefunction to undergo its standard unitary evolution over the whole course of history, and then, at the end of time, we imagine a single measurement being performed on the final state to determine the actual content of reality.  In Kent's words, `\emph{an event occurs if and only if it leaves effective records in the final time ... measurement}' \cite{Kent}. So for example in ref \cite{2017Kent} Kent imagines a final measurement on the positions of photons which have been reflected off matter at various points, and then  defines the beables at $x$ by the expectation value of some operators (e.g. the stress-energy tensor components) at  $x$ conditional  on the detections of photos outside the future lightcone of $x$.   

In Kent’s picture, all observed events are `absolute' but there are a few cases where one might expect an event to be observed but as a matter of fact no event gets observed. In general,  Kent's approach entails that macroscopic events will reliably occur as expected, since decoherence encodes these events robustly in the state of their environment and thus they will be actualised by the final measurement. And of course `observed events' are always macroscopic events - for even if the event itself is not macroscopic, the observer and the records made in their brain are macroscopic - so `observed events' are typically actualised, which is to say they actually occur and are actually observed. But in a special case where all records of a macroscopic events are subsequently destroyed, that macroscopic event will \emph{not} be actualised by the final measurement; so if the macroscopic event is an observed event, that event will not occur and will not be observed. And this is exactly what happens in the  non-absoluteness theorems -  for by definition, a supermeasurement can be performed on an observer only if decoherence is controlled and all records of their previous observations are completely erased by the supermeasurement. So the non-absoluteness theorems pertain to exactly the kinds of special cases in which Kent's model says certain observations do not in fact occur. For example, in the Bong et al experiment, whenever Alice performs a supermeasurement on Chidi that  supermeasurement destroys all the records of Chidi's  measurement,  so according to Kent's approach Chidi's measurement outcome is simply not an `observed event' in this instance, and therefore absoluteness of observed events does not entail that it must have a unique, well-defined outcome. 

 As with RQM+CPL, we can think of Kent's approach as postulating a divergence between dynamical states and intrinsic conditions. For example, in the Bong et al experiment, Kent's approach predicts that Alice's measurement outcomes will always be as if Chidi is in the state $\psi$, which is to say the dynamical state of Chidi relative to Alice is $\psi$. But we also know that if Alice in fact chooses to perform a supermeasurement,  Chidi's measurement is not actualised by the final measurement, so  Chidi may not have an intrinsic condition at all during this time, or at least, his intrinsic condition will not be at similar to the intrinsic condition normally associated with  a classical observer having a conscious experience -   in a sense  he doesn't really exist at all during this time. Yet Alice is able to dynamically access his degrees of freedom and perform measurements on him even though he `does not exist,' thus exhibiting a dramatic divergence between his intrinsic condition and his dynamical state. Thus  Kent's approach exhibits a version of Type-II  disaccord in the context of the Bong et al experiment: if Alice assumes that the fact that Chidi has dynamical state $\psi$ relative to her means that he is having some kind of indefinite `superposed' experiences, or at least that he is having \emph{some} experiences, then her inference will fail to match Chidi's experiences, since actually he is not having any experiences at this time.

 However, Kent's approach does not exhibit Type-III disaccord: if Alice asks Chidi about his measurement outcome, then the conversation itself forms a part of the records of his measurement outcome, and by construction the records arriving at the final state must match the event which is actualised by the final measurement, so what Alice hears Chidi saying will indeed match what he originally perceived. And note that quantum  mechanics is universal in the first-person sense in Kent's view, because it entails that no observer will ever observe anything that is incompatible with unitary quantum mechanics, although there are some cases where this is achieved only in virtue of certain observers failing to observe anything at all. Thus we can conclude that Kent's approach must violate one of the other assumptions of the Bong et al theorem; and indeed, it evidently exhibits the subtle kind of retrocausality discussed in section \ref{retro}, since we have seen that the result of Chidi’s measurement depends on what Alice chooses to measure. Note however that while in RQM+CPL Chidi may get two different results depending on what Alice chooses to measure, in Kent’s picture it is instead the case that if Alice chooses to ask Chidi about his outcome Chidi gets some result, whereas if Alice performs a supermeasurement Chidi does not get any result at all. So although both RQM+CPL and Kent's approach make use of relational dynamical states and retrocausality, they implement this in quite different ways\footnote{One might worry that Kent's approach has avoided the epistemic problems associated with Type-III disaccord only by introducing another kind of serious epistemic problem, since it entails that some of our beliefs about the past are very seriously mistaken - sometimes we believe an event has occurred when in fact it did not occur at all. This is a significant revision of our usual view of the past, and one might think it could undermine our scientific practice of using records of the past to empirically confirm theories. However, note that in Kent’s picture, the only kinds of events which fail to occur are events such that there are no ongoing records of the relevant event; so in Kent’s picture, we may rest assured that whenever we \emph{do} have access to a record of an event, that event most likely did occur in the way suggested by the record, and if it did not the reason is just an ordinary one like human error. So even though in Kent’s picture it is occasionally true that events don’t occur when we expect them to, nonetheless we can still expect records of past events to be generally reliable and thus we can make use of them for scientific confirmation.}.

\section{Ormrod and Barrett \label{OB}}

We have deferred discussion of the Ormrod and Barrett result to this final section because it is importantly different to the other cases. For the Wigner's Friend scenario and the Bong et al and Lawrence et al theorems, we were able to show that even if MOPO approaches are ruled out, and even if we are determined to maintain Locality and No-Superdeterminism, what is demonstrated by these theorems is only the existence of disaccord, not necessarily any kind of metaphysically radical non-absoluteness; therefore we were able to suggest a way of responding to these theorems in a way which maintains Locality, No-Superdeterminism, the existence of only one outcome per observer and the universality of unitary quantum mechanics in the first-person sense without any metaphysically radical non-absoluteness,  by simply relativizing the dynamical state rather than the observed events. 

But no such option exists for the Ormrod and Barrett case: in this experiment, FIQT predicts that there is a non-zero probability that Alice and Bob get outcomes such that it is mathematically impossible that   all observers  have exactly one non-relativized outcome and all of these unique outcomes obey FIQT for any choice of spacelike slice. Therefore on runs of the experiment when Alice and Bob do get this outcome, it follows that if all four observers have a unique non-relative outcome, on at least one spacelike slice there must be a pair of outcomes which is predicted by FIQT applied on that slice to be impossible; so FIQT cannot be universally correct if all measurement outcomes are unique and non-relativized.   Relativizing the dynamical state doesn't help here, since the theorem doesn't require any observers to communicate their results to one another and thus it doesn't make any assumptions about the relation between the outcomes of dynamical interactions with an observer and the actual experiences of that observer. Adopting a picture like Kent's does help in a certain sense - it would resolve the apparent contradiction by simply saying that the measurements of Chidi and Divya don't actually take place, since all records of them are subsequently erased, so we don't get outcomes predicted to be impossible by FIQT on the Chidi-Divya spacelike slice because we don't get outcomes at all on that slice. But presumably Ormrod and Barrett would take the view that such an approach does not fully uphold their assumption of the universality of FIQT, since it entails that some outcomes predicted by FIQT simply don't occur at all. 

Nor can we straightforwardly argue that what really follows from the assumptions of the theorem is disaccord rather than non-absoluteness. We \emph{can} obtain Type-I disaccord from this scenario, but only by taking a deflationary appraoch to FIQT and regarding it as a description, not of the actual  measurement outcomes that occur, but of some inferences that could reasonably be made by observers in a scenario of this kind. So for example, suppose Alice and Bob  compare their results and then make   an inference about Divya's measurement result by imagining a collapse on the Alice-Divya spacelike slice and using what they know about Alice's outcome, and  an inference about Chidi's measurement result by imagining a collapse on the Bob-Chidi spacelike slice and using what they know about Bob's outcome, and likewise, Chidi makes an inference about Divya's measurement result by imagining a collapse on the Chidi-Divya spacelike slice and using his knowledge of his own outcome, and Divya  makes an inference about Chidi's result in a similar way. Then the import of the Ormrod and Barrett theorem is that there exists a possible set of measurement outcomes for Alice and Bob such that it is mathematically guaranteed that not all of these inferences made in accordance with FIQT can be correct. Thus we arrive at Type-I disaccord, because at least one person in this scenario must end up making a wrong inference about someone else's outcome, even though they applied quantum mechanics correctly. But evidently this approach does not really maintain the universality of unitary quantum mechanics in the FIQT sense, since it involves accepting that on at least one spacelike slice the predictions of FIQT are not right. 
 
So the theorem of Ormrod and Barrett seems like it really could provide an argument  for   non-absoluteness in a metaphysically radical sense: if quantum mechanics is truly universal in the FIQT sense, and if MOPO approaches have been ruled out, then the only remaining option is that there must be some kind of radical relativization going on. However,   it is important to note that this follows only because of the stronger universality assumption made by Ormrod and Barrett - no such result can be derived using purely first-person universality. And therefore  the kind of `non-absoluteness'   involved here cannot be the kind of non-absoluteness that features in ordinary relational and perspectival approaches to quantum mechanics,  for approaches of this kind typically postulate only first-person universality, and indeed they often forbid comparisons between measurement results obtained by different observers. Thus in such approaches it is simply not meaningful to apply quantum mechanics to the results of two different observers on the same spacelike slice in the way that Ormrod and Barrett do, unless of course those observers later interact in some way which makes the outcome of one physically relevant to the outcome of the other, which is not the case in the Ormrod and Barrett scenario. Thus the very definition of FIQT is actually incompatible with many relational and perspectival interpretations, so it should  be kept in mind that the assumptions going into the Ormrod and Barrett non-absoluteness theorem are   incompatible with a number of the metaphysically radical interpretations that are sometimes argued to be supported by the non-absoluteness theorems. 

Indeed, it seems quite hard to understand what  kind of relational or perspectival approach \emph{could} help with the problem posed by the theorem of Ormrod and Barrett. Usually in relational and perspectival approaches we are called upon   to make reference to  `Divya's measurement outcome relative to Alice' only in cases where Alice physically measures some variable encoding Divya's outcome, or where Divya's outcome is correlated with or has consequences for something which Alice can physically interact with after the experiment has concluded; the existence of  real interactions of this kind establishing real physical relations between the observers is what endows the notion of `Divya's measurement outcome relative to Alice' with  physical content.  But no such interaction can possibly  occur in the Ormrod and Barrett scenario, since Divya's measurement outcome is erased before it can have any consequences for Alice, so it is not clear what physical significance  claims about `Divya's measurement outcome relative to Alice'  could possibly have in this context - if this relativized outcome is not relevant to any of Alice's future predictions or interactions, and it is not a fact about Divya's experiences or Alice's experiences, what \emph{is} it a fact about? Perhaps one might think that `the outcome of Divya relative to Alice' in this context has \emph{epistemic} rather than physical significance:  Divya's measurement outcome `has a value’ relative to Alice in the sense that Alice believes something about it, or it would be rational for her to believe something about it based on her understanding of quantum mechanics. But as we have already noted, this epistemic approach would surely not satisfy anyone who is determined that quantum mechanics must be universal in the FIQT sense, since it just amounts to saying that inferences made using FIQT may be reasonable but they are nonetheless sometimes wrong. 

Thus we are not convinced that there is any coherent version of metaphysically radical non-absoluteness which can be supported by the Ormrod and Barrett theorem. Rather, the theorem is probably best understood as an argument in support of  a MOPO approach, such as the Everett interpretation or perhaps a consistent-history style approach. And if MOPO approaches were to be ruled out, then perhaps the appropriate conclusion to draw from the   theorem would simply be that quantum mechanics cannot possibly be universal in the FIQT sense -  if there is no coherent version of `non-absoluteness' which would allow us to reconcile the universality of quantum mechanics in the FIQT sense with the stipulation that all observers always see a single  outcome to a given measurement, then for those who believe that all observers should always see a single  outcome,  the theorem becomes a reductio ad absurdum against FIQT. 

In this connection, it is important to note that although FIQT is motivated by the desire to have compatibility with relativity, it is certainly possible to have relativistically covariant approaches which exhibit universality only in the weaker first-person sense. For example, Kent's approach  is explicitly designed to be Lorentz-covariant, and yet it has the consequence that quantum mechanics is universal only in the first-person and not in the FIQT sense, since it tells us that some of the events predicted by FIQT do not actually occur. Similarly, RQM+CPL   appears to be  relativistically covariant when formulated as an `all-at-once' theory, but it tells us that quantum mechanics is universal only in the first-person and not in the FIQT sense, since it does not require that quantum mechanics makes correct predictions when applied across the perspectives of two observers who never have an opportunity to share their results. So the claim made by Ormrod and Barrett, `\emph{Given a version of quantum theory that models measurements unitarily and which fits naturally with special relativity, it simply cannot be true that there are absolute observed outcomes}' is not completely right: their vision of what it might look like for a version of quantum theory to `fit naturally with special relativity' is  too narrow, as it fails to take into account the possibility of models in which there are fewer observed events than we might naturally imagine, or all-at-once models with dynamically relational states, or other novel possibilities that have not yet even been thought of. So in our view it is reasonable   to respond to the Ormrod and Barrett theorem by simply rejecting FIQT, although of course  the MOPO route still remains open.

Additionally, it is very interesting   that the universality of quantum mechanics in the first-person sense appears to be perfectly compatible with the existence of only one outcome per observer \emph{and} the absence of any metaphysically radical non-absoluteness - only a much stronger universality assumption  could compel us to accept either a MOPO approach or some kind of metaphysically radical non-absoluteness.   Since universality in the first-person sense is naturally associated with relational and perspectival views, this suggests that the strong metaphysical claims often tied to  such views may be a  misunderstanding of what quantum mechanics is really telling us about relationality - such approaches should all along  have been understood as being \emph{dynamically} rather than  metaphysically relational. Additionally, since we can only really have direct empirical evidence for universality in the first-person sense and not in the FIQT sense, this conclusion may perhaps be regarded as an  argument in favour of the absoluteness of observed events: our immediate empirical evidence seems to be telling us that the world has relational features but that it stays strictly within the limits of relationality  compatible with having one outcome per observer without any metaphysically radical non-absoluteness, and one natural explanation for that would be that observed events are, in fact, absolute!
 
\section{Conclusion}

In this article we have reached a somewhat negative conclusion about the idea of `non-absoluteness of observed events.’ That is, we agree that the Everett interpretation and other MOPO appraoches would genuinely involve something that could be referred to as non-absoluteness of observed events, but we are not convinced that the Wigner's Friend scenario or the non-absoluteness theorems offer compelling evidence for any alternative approach involving metaphysically radical non-absoluteness. For we have seen that  the Wigner's Friend scenario and the Bong et al and Lawrence et al theorems can be understood simply as  demonstrating that if unitary quantum mecahnics is universal and certain auxiliary assumptions hold, then there must exist instances of Type-II and/or Type-III disaccord,  and we have argued that these kinds of disaccord do not need to be understood in terms of metaphysically radical non-absoluteness, since we always have the option of relativizing the \emph{dynamical} state whilst maintaining the absoluteness of observed events. Of course Type-III disaccord is certainly unappealing in many ways and we agree that it should be avoided, but postulating metaphysically radical non-absoluteness does not help at all with this problem and indeed makes things significantly worse. Meanwhile, the theorem of Ormrod and Barrett does seem like it could offer a real argument for metaphysically radical non-absoluteness, but it only achieves this by making an unusually strong universality assumption, and it is unclear that accepting metaphysically radical non-absoluteness is a better option than simply rejecting this strong universality assumption (or alternatively, adopting a MOPO approach). 

Now, we suspect that some of our negative comments about metaphysically radical non-absoluteness may invite a response of the form: `You are prejudiced against this kind of picture because you are too attached to your  naive classical worldview and are thus incapable of properly comprehending a radically relational/perspectival approach.' And this may be so. But our central point in this article is simply that these kinds of radical approaches are not \emph{necessary}; and given that they appear to lead to severe problems for the epistemology of science, we probably should not adopt them unless there is genuinely no other choice. Moreover, it seems  that relativizing the dynamical state while maintaining `absolute' intrinsic conditions and observed events gives us nearly everything we might hope to gain from a relational or perspectival approach but without  the associated dangers for scientific rationality, so we believe that the possibility of developing views which are relational in a less radical way deserves more attention than it has so far received.

Finally, although we don’t think the non-absoluteness theorems should be interpreted as demonstrating the non-absoluteness of observed events in a metaphysically radical sense, we do think they  provide important new insights into  quantum mechanics. For example, the scenarios and theorems discussed in this paper together both make a clear case for dynamically relational approaches and also give  a sense of the shape which a dynamically relational approach would have to take -  in particular,   the   results of Bong et al and Lawrence et al suggest that if we want to avoid Type-III disaccord  whilst also continuing to maintain the universality of unitary quantum mechanics in the first person sense and the existence of only one outcome per observer, we may be forced to compromise by allowing the violation of Locality, and/or allowing some superdeterminism or retrocausality.  So in our view, these recent results demonstrate that the common prejudice against the measurement problem as unscientific and unsolvable is unfounded: the combination of these non-absoluteness theorems, the  need to avoid epistemic irrationality, and the mandate to successfully reproduce the predictions of QFT is now forcing us down an increasingly narrow bottleneck, so progress is certainly being made and there is cause for optimism that we may eventually converge on a fully acceptable solution.

\section{Acknowledgements}

Thanks to Nick Ormrod for reading and discussing a draft of this paper, and thanks to Howard Wiseman and Yil\'{e} Ying  for very helpful conversations on these topics.   This publication was made possible in part through the support of the ID \#62312 grant from the John Templeton Foundation, as part of the project \href{https://www.templeton.org/grant/the-quantum-information-structure-of-spacetime-qiss-second-phase}{‘The Quantum Information Structure of Spacetime’ (QISS)}. The opinions expressed in this publication are those of the author and do not necessarily reflect the views of the John Templeton Foundation.


\begin{thebibliography}{10}

\bibitem{2022NatRP...4..628B}
{\v{C}}aslav {Brukner}.
\newblock {Wigner's friend and relational objectivity}.
\newblock {\em Nature Reviews Physics}, 4(10):628--630, September 2022.

\bibitem{mbitbol}
Michel Bitbol.
\newblock Is the life-world reduction sufficient in quantum physics?
\newblock {\em Continental Philosophy Review}, 54, 12 2021.

\bibitem{subjectivefacts}
Alessandro Fedrizzi and Massimiliano Proietti.
\newblock {Quantum physics: our study suggests objective reality doesn’t
  exist}, Accessed August 2023.

\bibitem{Oldofredi2022-OLDTRD}
Andrea Oldofredi.
\newblock The relational dissolution of the quantum measurement problems.
\newblock {\em Foundations of Physics}, 53(1):1--24, 2022.

\bibitem{Pienaar_2021}
Jacques Pienaar.
\newblock {QBism} and relational quantum mechanics compared.
\newblock {\em Foundations of Physics}, 51(5), oct 2021.

\bibitem{wigner1995remarks}
Eugene~P Wigner.
\newblock Remarks on the mind-body question.
\newblock In {\em Philosophical reflections and syntheses}, pages 247--260.
  Springer, 1995.

\bibitem{Bong_2020}
Kok-Wei Bong, An{\'{\i}}bal Utreras-Alarc{\'{o}}n, Farzad Ghafari, Yeong-Cherng
  Liang, Nora Tischler, Eric~G. Cavalcanti, Geoff~J. Pryde, and Howard~M.
  Wiseman.
\newblock A strong no-go theorem on the wigner's friend paradox.
\newblock {\em Nature Physics}, 16(12):1199--1205, aug 2020.

\bibitem{Lawrence_2023}
Jay Lawrence, Marcin Markiewicz, and Marek {\.{Z} }ukowski.
\newblock Relative facts of relational quantum mechanics are incompatible with
  quantum mechanics.
\newblock {\em Quantum}, 7:1015, may 2023.

\bibitem{ormrod2022nogo}
Nick Ormrod and Jonathan Barrett.
\newblock A no-go theorem for absolute observed events without inequalities or
  modal logic, 2022.

\bibitem{schmid2023review}
David Schmid, Yìlè Yīng, and Matthew Leifer.
\newblock A review and analysis of six extended wigner's friend arguments,
  2023.

\bibitem{wiseman2023thoughtful}
Howard~M. Wiseman, Eric~G. Cavalcanti, and Eleanor~G. Rieffel.
\newblock A "thoughtful" local friendliness no-go theorem: a prospective
  experiment with new assumptions to suit, 2023.

\bibitem{FuchsMermin}
C.~A. {Fuchs}, N.~D. {Mermin}, and R.~{Schack}.
\newblock {An introduction to QBism with an application to the locality of
  quantum mechanics}.
\newblock {\em American Journal of Physics}, 82:749--754, August 2014.

\bibitem{1996cr}
Carlo Rovelli.
\newblock Relational quantum mechanics.
\newblock {\em International Journal of Theoretical Physics},
  35(8):1637–1678, Aug 1996.

\bibitem{Frigg2009}
Roman Frigg.
\newblock {\em {GRW Theory (Ghirardi, Rimini, Weber Model of Quantum
  Mechanics)}}, pages 266--270.
\newblock Springer Berlin Heidelberg, Berlin, Heidelberg, 2009.

\bibitem{Tumulka2006}
R.~{Tumulka}.
\newblock {A Relativistic Version of the Ghirardi Rimini Weber Model}.
\newblock {\em Journal of Statistical Physics}, 125:821--840, November 2006.

\bibitem{london1939theorie}
Fritz London, Edmond Bauer, and Paul Langevin.
\newblock La th{\'e}orie de l'observation en m{\'e}canique quantique.
\newblock {\em (No Title)}, 1939.

\bibitem{https://doi.org/10.48550/arxiv.2205.00568}
David Wallace.
\newblock The sky is blue, and other reasons quantum mechanics is not
  underdetermined by evidence, 2022.

\bibitem{AdlamEverett}
Emily Adlam.
\newblock {The Problem of Confirmation in the Everett Interpretation}.
\newblock {\em Studies in History and Philosophy of Science Part B: Studies in
  History and Philosophy of Modern Physics}, 47:21 -- 32, 2014.

\bibitem{https://doi.org/10.48550/arxiv.2203.16278}
Emily Adlam.
\newblock {Does Science need Intersubjectivity? The Problem of Confirmation in
  Orthodox Interpretations of Quantum Mechanics}.
\newblock {\em Synthese}, 2022.

\bibitem{Greaves}
H.~{Greaves}.
\newblock {Understanding Deutsch's probability in a deterministic multiverse}.
\newblock {\em eprint arXiv:quant-ph/0312136}, December 2003.

\bibitem{Wallace}
David Wallace.
\newblock Everett and structure.
\newblock {\em Studies in History and Philosophy of Science Part B: Studies in
  History and Philosophy of Modern Physics}, 34(1):87--105, 2003.

\bibitem{vaidman1996schizophrenic}
Lev Vaidman.
\newblock On schizophrenic experiences of the neutron or why we should believe
  in the many-worlds interpretation of quantum theory, 1996.

\bibitem{Bohr1987-BOHTPW}
Niels Bohr.
\newblock {\em The Philosophical Writings of Niels Bohr}.
\newblock Ox Bow Press, 1987.

\bibitem{Heisenberg1958-HEIPAP}
Werner Heisenberg.
\newblock {\em Physics and Philosophy: The Revolution in Modern Science}.
\newblock New York: Harper, 1958.

\bibitem{pauli1994writings}
Wolfgang Pauli, Charles~P Enz, and Karl von Meyenn.
\newblock {\em Writings on physics and philosophy}.
\newblock Springer, 1994.

\bibitem{Zeilinger1999-ZEIAFP}
Anton Zeilinger.
\newblock A foundational principle for quantum mechanics.
\newblock {\em Foundations of Physics}, 29(4):631--643, 1999.

\bibitem{Zeilinger2002}
Anton Zeilinger.
\newblock {\em Bell's Theorem, Information and Quantum Physics}, pages
  241--254.
\newblock Springer Berlin Heidelberg, Berlin, Heidelberg, 2002.

\bibitem{brukner2015quantum}
Caslav Brukner.
\newblock On the quantum measurement problem.
\newblock 2015.

\bibitem{articlebanana}
Jeffrey Bub.
\newblock Bananaworld: Quantum mechanics for primates.
\newblock 11 2012.

\bibitem{demopoulos2012generalized}
William Demopoulos.
\newblock Generalized probability measures and the framework of effects.
\newblock In {\em Probability in Physics}, pages 201--217. Springer, 2012.

\bibitem{Janas2021-JANUQR}
Michael Janas, Michael~E. Cuffaro, and Michel Janssen.
\newblock {\em Understanding Quantum Raffles: Quantum Mechanics on an
  Informational Approach - Structure and Interpretation (Foreword by Jeffrey
  Bub)}.
\newblock Springer, 2021.

\bibitem{2004neoc}
Willem~M. de~Muynck.
\newblock {Towards a Neo-Copenhagen Interpretation of Quantum Mechanics}.
\newblock {\em Foundations of Physics}, 34(5):717–770, May 2004.

\bibitem{QBismintro}
C.~A. {Fuchs}, N.~D. {Mermin}, and R.~{Schack}.
\newblock {An introduction to QBism with an application to the locality of
  quantum mechanics}.
\newblock {\em American Journal of Physics}, 82:749--754, August 2014.

\bibitem{pittphilsci19664}
Andrea~Di Biagio and Carlo Rovelli.
\newblock {Relational Quantum Mechanics is about Facts, not States: A reply to
  Pienaar and Brukner}, October 2021.

\bibitem{Gisin2013}
M.~{Esfeld} and N.~{Gisin}.
\newblock {The GRW flash theory: a relativistic quantum ontology of matter in
  space-time?}
\newblock {\em ArXiv e-prints}, October 2013.

\bibitem{Moreno_2022}
George Moreno, Ranieri Nery, Cristhiano Duarte, and Rafael Chaves.
\newblock Events in quantum mechanics are maximally non-absolute.
\newblock {\em Quantum}, 6:785, aug 2022.

\bibitem{Price_1994}
Huw Price.
\newblock A neglected route to realism about quantum mechanics.
\newblock {\em Mind}, 103(411):303–336, 1994.

\bibitem{Price2005-PRICP}
Huw Price.
\newblock Causal perspectivalism.
\newblock In Huw Price and Richard Corry, editors, {\em Causation, Physics, and
  the Constitution of Reality: Russell's Republic Revisited}. Oxford University
  Press, 2005.

\bibitem{adlam2022roads}
Emily Adlam.
\newblock {Two Roads to Retrocausality}.
\newblock {\em Synthese}, 200(422), 2022.

\bibitem{Adlam2023-ADLITC-3}
Emily Adlam.
\newblock Is there causation in fundamental physics? new insights from process
  matrices and quantum causal modelling.
\newblock {\em Synthese}, 201(5):1--40, 2023.

\bibitem{dibiagio2021relational}
Andrea~Di Biagio and Carlo Rovelli.
\newblock Relational quantum mechanics is about facts, not states: A reply to
  pienaar and brukner, 2021.

\bibitem{https://doi.org/10.48550/arxiv.2203.13342}
Emily Adlam and Carlo Rovelli.
\newblock Information is physical: Cross-perspective links in relational
  quantum mechanics, 2022.

\bibitem{cavalcanti2023consistency}
Eric~G. Cavalcanti, Andrea~Di Biagio, and Carlo Rovelli.
\newblock On the consistency of relative facts, 2023.

\bibitem{Kent}
A.~{Kent}.
\newblock {Solution to the Lorentzian quantum reality problem}.
\newblock {\em Phys Rev A}, 90(1):012107, July 2014.

\bibitem{2015KentL}
Adrian Kent.
\newblock Lorentzian quantum reality: postulates and toy models.
\newblock {\em Philosophical Transactions of the Royal Society A: Mathematical,
  Physical and Engineering Sciences}, 373(2047):20140241, Aug 2015.

\bibitem{2017Kent}
Adrian Kent.
\newblock Quantum reality via late-time photodetection.
\newblock {\em Physical Review A}, 96(6), Dec 2017.

\end{thebibliography}
\end{document}